\documentclass[letterpaper, oneside, 12pt]{article}
\usepackage[round]{natbib}
\usepackage[reqno]{amsmath}
\usepackage{amsthm,amsfonts,amssymb}
\usepackage{graphicx}
\usepackage{amsfonts}
\usepackage{latexsym}
\usepackage{subfigure}
\usepackage{url}
\usepackage{algorithmic}\label{•}
\usepackage{longtable}
\usepackage{multirow}
\usepackage{fancyvrb}
\usepackage{float}
\usepackage{setspace}
\usepackage{slashbox}
\usepackage[hmargin=1.0in,vmargin=1in]{geometry}
\usepackage{setspace}
\usepackage{rotating}
\usepackage{enumerate}
\usepackage{comment}
\usepackage{booktabs}
\bibpunct{(}{)}{;}{a}{}{,}

\DefineVerbatimEnvironment{Code}{Verbatim}{fontsize=\small}
\newcommand{\ignore}[1]{}

\usepackage{lineno}

\newcommand{\bet}{ \mbox{\boldmath $ \eta $} }

\newcommand{\bbeta}{ \mbox{\boldmath $ \beta $} }

\newcommand{\balpha}{ \mbox{\boldmath $ \alpha $} }

\newcommand{\bepsilon}{ \mbox{\boldmath $\epsilon$}}

\newcommand{\btheta}{ \mbox{\boldmath $ \theta $} }

\newcommand{\bmu}{ \mbox{\boldmath $\mu$} }
\newcommand{\bOmega}{ \mbox{\boldmath $\Omega$} }

\newcommand{\bTheta}{ \mbox{\boldmath $\Theta$} }

\newcommand{\bLambda}{ \mbox{\boldmath $\Lambda$} }

\newcommand{\btau}{ \mbox{\boldmath $\tau$} }

\newcommand{\bzero}{\textbf{0}}
\newcommand{\bones}{\textbf{1}}

\newcommand{\bA}{\textbf{A}}
\newcommand{\bb}{\textbf{b}}
\newcommand{\bB}{\textbf{B}}
\newcommand{\bc}{\textbf{c}}
\newcommand{\bC}{\textbf{C}}

\newcommand{\bD}{\textbf{D}}
\newcommand{\be}{\textbf{e}}

\newcommand{\bK}{\textbf{K}}

\newcommand{\bH}{\textbf{H}}

\newcommand{\bI}{\textbf{I}}
\newcommand{\bg}{\textbf{g}}
\newcommand{\bG}{\textbf{G}}
\newcommand{\bJ}{\textbf{J}}
\newcommand{\bL}{\textbf{L}}
\newcommand{\bm}{\textbf{m}}
\newcommand{\bM}{\textbf{M}}
\newcommand{\bn}{\textbf{N}}

\newcommand{\bO}{\textbf{O}}

\newcommand{\bq}{\textbf{q}}
\newcommand{\bQ}{\textbf{Q}}

\newcommand{\bs}{\textbf{s}}

\newcommand{\bT}{\textbf{T}}
\newcommand{\bu}{\textbf{u}}

\newcommand{\bv}{\textbf{v}}
\newcommand{\bV}{\textbf{V}}
\newcommand{\bw}{\textbf{w}}
\newcommand{\bW}{\textbf{W}}
\newcommand{\bx}{\textbf{x}}
\newcommand{\bX}{\textbf{X}}
\newcommand{\by}{\textbf{y}}
\newcommand{\bY}{\textbf{Y}}
\newcommand{\bZ}{\textbf{Z}}
\newcommand{\bz}{\textbf{z}}

\newcommand{\T}{\mathrm{T}}

\newcommand{\tildeu}{\tilde{u}}
\newcommand{\tildev}{\tilde{v}}

\newcommand{\tildeeps}{\tilde{\epsilon}}

\newcommand{\given}{\,|\,}

\newcommand{\E}{\mathbb E}

\newcommand{\s}{\mathbf s}

\newcommand{\taus}{\tau^2}

\newcommand{\calD}{{\cal D}}

\newcommand{\calH}{{\cal H}}

\newcommand{\calL}{{\cal L}}
\newcommand{\calS}{{\cal S}}

\newcommand{\calX}{{\cal X}}

\usepackage[printwatermark]{xwatermark}
\usepackage{xcolor}
\usepackage{graphicx}

\begin{document}

\title{\vspace{-2.0cm}Joint hierarchical models for sparsely sampled high-dimensional LiDAR and forest variables}

\author{Andrew O. Finley, Sudipto Banerjee, Yuzhen Zhou,\\Bruce D. Cook, and Chad Babcock\let\thefootnote\relax\footnote{Andrew O. Finley is Associate Professor, Departments of Forestry and Geography, Michigan State University, East Lansing, MI 48824.}
}
\date{}
\maketitle
\vspace{-0.5in}
\begin{abstract}
Recent advancements in remote sensing technology, specifically Light Detection and Ranging (LiDAR) sensors, provide the data needed to quantify forest characteristics at a fine spatial resolution over large geographic domains. From an inferential standpoint, there is interest in prediction and interpolation of the often sparsely sampled and spatially misaligned LiDAR signals and forest variables. We propose a fully process-based Bayesian hierarchical model for above ground biomass (AGB) and LiDAR signals. The process-based framework offers richness in inferential capabilities, e.g., inference on the entire underlying processes instead of estimates only at pre-specified points. Key challenges we obviate include misalignment between the AGB observations and LiDAR signals and the high-dimensionality in the model emerging from LiDAR signals in conjunction with the large number of spatial locations. We offer simulation experiments to evaluate our proposed models and also apply them to a challenging dataset comprising LiDAR and spatially coinciding forest inventory variables collected on the Penobscot Experimental Forest (PEF), Maine. Our key substantive contributions include AGB data products with associated measures of uncertainty for the PEF and, more broadly, a methodology that should find use in a variety of current and upcoming forest variable mapping efforts using sparsely sampled remotely sensed high-dimensional data.  
\end{abstract}


\section{Introduction}\label{Sec: Intro}
Coupling forest inventory with remotely sensed Light Detection and Ranging (LiDAR) datasets using regression models offers an attractive approach to mapping forest variables at stand, regional, continental, and global scales. LiDAR data have shown great potential for use in estimating spatially explicit forest variables over a range of geographic scales \citep{asner2009, babcock2013, finley2011, iqbal2013, muss2011, naesset2011, neigh2013}. Encouraging results from these and many other studies have spurred massive investment in new LiDAR sensors and sensor platforms, as well as extensive campaigns to collect field-based calibration data.

Much of the interest in LiDAR based forest variable mapping is to support carbon monitoring, reporting, and verification (MRV) systems, such as defined by the United Nations Programme on Reducing Emissions from Deforestation and Forest Degradation (UN-REDD) and NASA's Carbon Monitoring System (CMS) \citep{le2011, ometto2014, REDD2009, CMS2010}. In these, and similar initiatives, AGB is the forest variable of interest because it provides a nearly direct measure of forest carbon (i.e., carbon comprises $\sim$50\% of wood biomass, \citealt{west2004}). Most efforts to quantify and/or manage forest ecosystem services, e.g., carbon, biodiversity, water, seek high spatial resolution wall-to-wall data products such as gridded maps with associated measures of uncertainty, e.g., point and associated credible intervals (CIs) at the pixel level. In fact several high profile international initiatives include language concerning the level of spatially explicit acceptable error in total forest carbon estimates, see, e.g., \cite{REDD2009} and \cite{UNFCCC2015}.

Many current LiDAR data acquisition campaigns focus on achieving complete coverage at a high spatial resolution over the domain of interest, e.g., resulting in a fine grid with each pixel yielding a high-dimensional LiDAR signal. In practice, a variety of non-statistical approaches are then used to characterize the LiDAR signals---effectively a dimension reduction step, \citet{Anderson2008}, \citet{Gonzalez2010}, \citet{muss2011}, \citet{Tonolli2011}, \citet{Popescu2008}, and \citet{babcock2013}. These signal characteristics serve as regressors in models where the outcome forest variables are measured at a relatively small set of georeferenced forest inventory plots. The regression model is then used to predict the forest outcome variables at all LiDAR pixels across the domain. This approach works well for small-scale forest variable mapping efforts. However, next generation LiDAR acquisition campaigns aimed at mapping and quantifying variables over large spatial extents, such as ICESat-2 \citep{abdalati2010, ICESAT2}, Global Ecosystem Dynamics Investigation LiDAR (GEDI) \citep{GEDI2014}, and NASA Goddard's LiDAR, Hyper-spectral, and Thermal (G-LiHT) imager \citep{cook2013,wired14}, will collect LiDAR data \emph{samples} from the domain of interest, e.g., using transect or cluster designs. The designs specify point-referenced LiDAR sampling across the domain extent and also over forest inventory plot locations (again for regression model calibration). In such settings the primary objective is still delivery of high resolution wall-to-wall predictive maps of forest variables, but also corresponding maps of LiDAR signal predictions at non-sampled locations. Further, to inform future LiDAR collection sampling designs, there is interest in characterizing the spatial dependence of within and, more importantly, among LiDAR signals. This information can help guide LiDAR sampling strategies with the aim to maximize some information gain criterion; see, e.g., \cite{xia06}, \cite{Mateu2014}. 

We propose a flexible framework to jointly model spatially misaligned LiDAR signals and forest inventory plot outcomes (e.g., AGB) that will $i$) automatically (i.e., no explicit variable selection step) extract information from the high-dimensional LiDAR signals to explain variability in the forest variable of interest, $ii$) estimate spatial dependence among and within LiDAR signals to improve inference and possibility help inform future LiDAR sampling strategies, and $iii$) provide full posterior predictive inference for both LiDAR signals and forest variables at locations where either one or neither of the data sources are available (i.e., wall-to-wall prediction). 

Meeting these objectives is particularly challenging for several reasons. From a computational standpoint each LiDAR signal is high-dimensional and the signals as well as the forest inventory plots are observed at a potentially large number of locations. From a model specification standpoint there are several sources of dependence that should be accommodated, including $i$) within and between LiDAR signals, $ii$) between LiDAR signals and spatially proximate forest variable measurements, and $iii$) residual spatial dependence in the signals and forest variables. These dependencies often result from strong vertical and horizontal similarities in forest structure caused by past management and/or natural disturbances. 

Our primary methodological contribution is the development of a modeling framework for high-dimensional misaligned data. Given the rich inference we seek (see preceding paragraph), our Bayesian hierarchical framework jointly models LiDAR signals and forest variables as a random process using latent Gaussian processes (GPs). This considerably enhances the computational burden of fitting them to datasets with a large number of spatial locations. The costs are exacerbated further by even a modest number of heights at which the LiDAR signal is observed. We achieve dimension reduction through bias-adjusted reduced-rank representations of the joint LiDAR-AGB process.      

The manuscript is organized as follows. Section~\ref{Sec: Data} provides an overview of the motivating dataset that comprises G-LiHT LiDAR and AGB measured at forest inventory plots on the Penobscot Experimental Forest (PEF) in Bradley, Maine. Section~\ref{Sec: Models} describes the proposed hierarchical model for the joint LiDAR-AGB process. The details on Bayesian prediction and implementation are given in the Supplemental Material. Section~\ref{Sec: illlustrations} offers an analysis of a synthetic dataset and PEF analysis. Finally, Section~\ref{Sec: Summary} concludes the manuscript with a brief summary and pointers toward future work.

\section{Data}\label{Sec: Data}
\begin{figure}[!htbp]
\centering
\subfigure[]{\includegraphics[width=8cm]{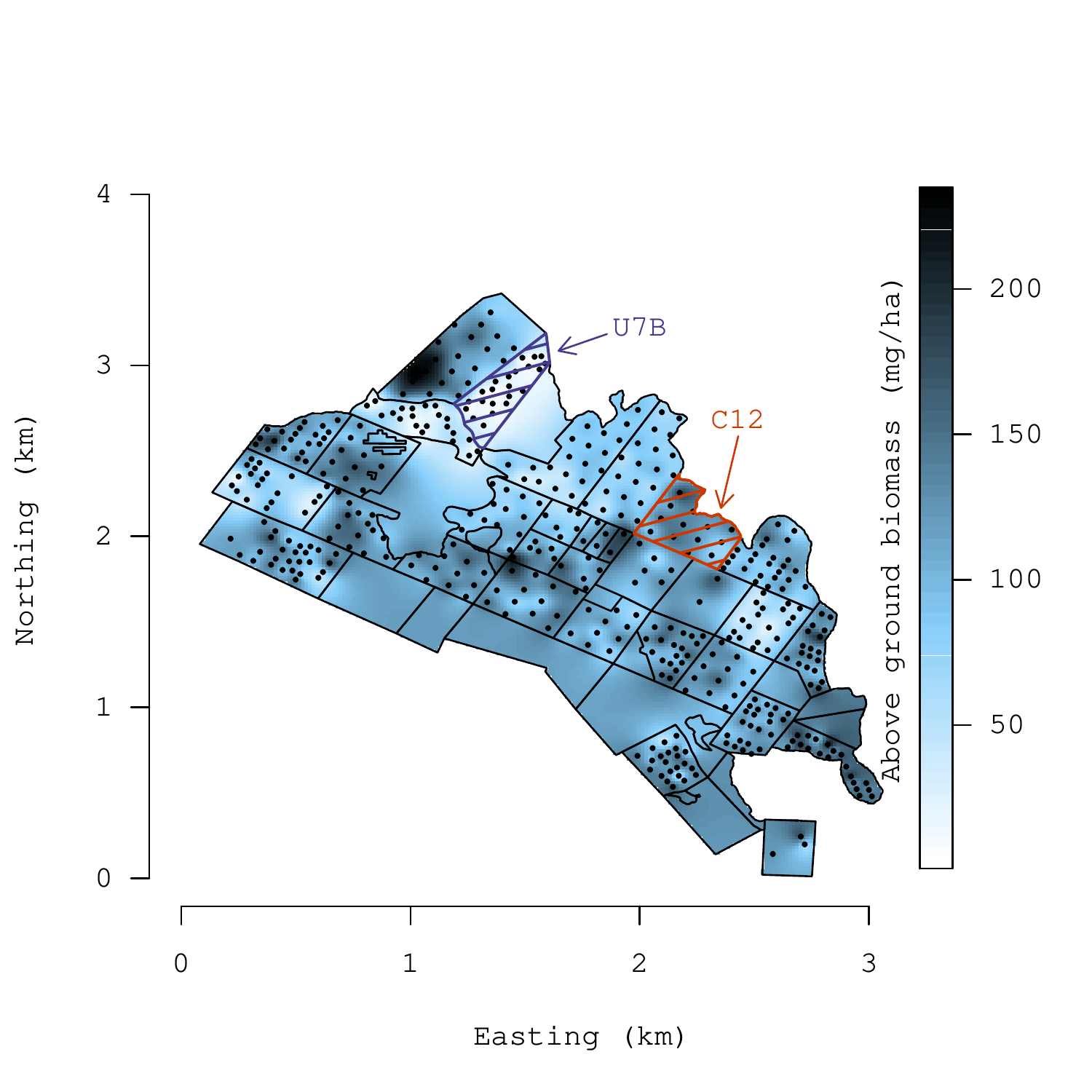}\label{pefMap}}
\subfigure[]{\includegraphics[width=8cm]{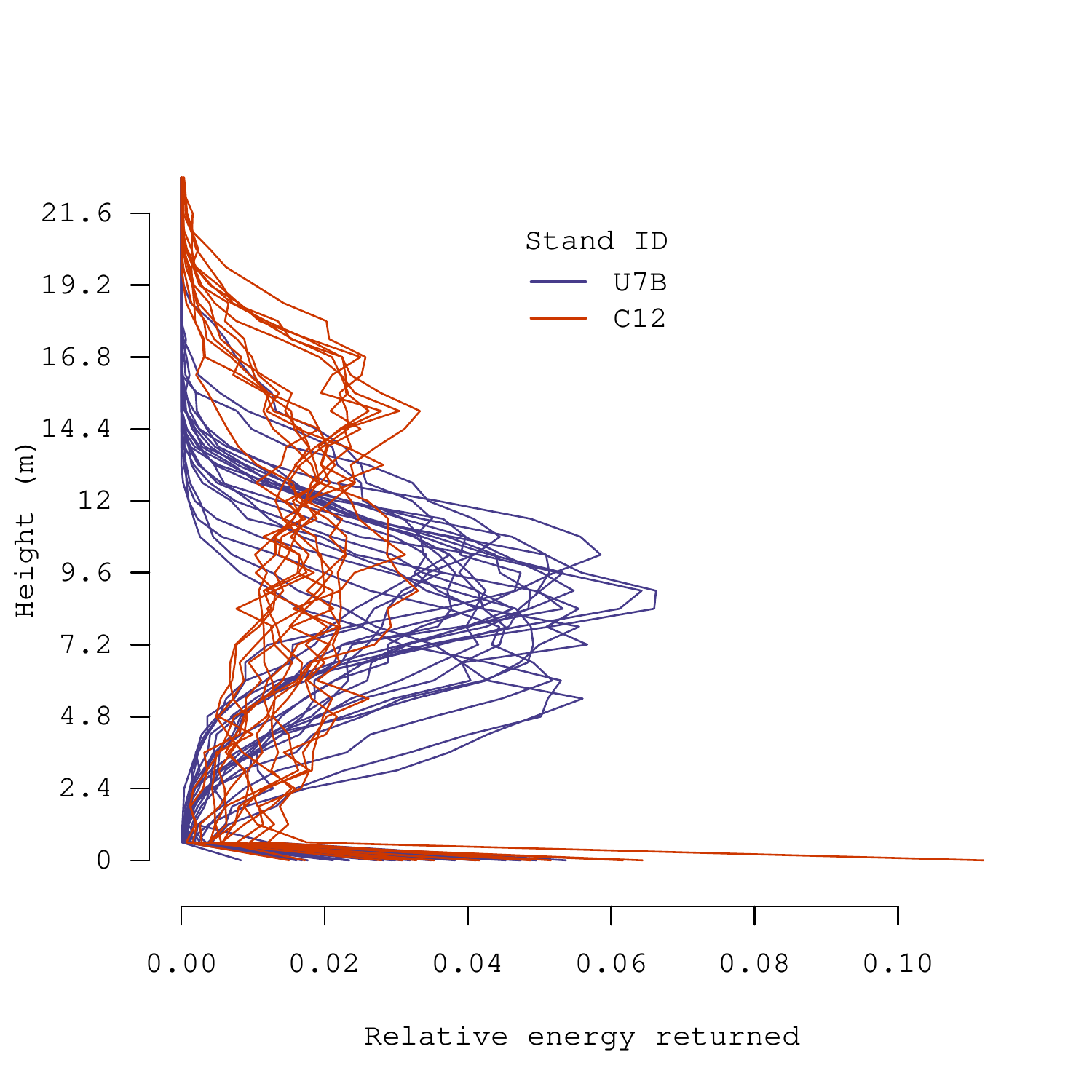}\label{pefLidar}}
\caption{\subref{pefMap} Penobscot Experimental Forest, Maine, with management units and forest inventory plot locations delineated as polygons and points, respectively. \subref{pefLidar} G-LiHT LiDAR signals observed at forest inventory plots highlighted in \subref{pefMap}.}\label{pef}
\end{figure}

The PEF is a 1600 ha tract of Acadian forest located in Bradley, Maine (44$^\circ$ 52' N, 68$^\circ$ 38' W). The forest is divided into over 50 management units (MU)---delineated as black polygons in Figure~\ref{pefMap}---that received management and monitoring since the 1950s \citep{sendak2003}. Within each MU, different silvicultural treatments are implemented, e.g., unregulated harvest, shelterwood, diameter limit cutting, or natural regeneration. Following procedures described in \cite{finley2014}, AGB (Mg/ha) was calculated for each of 451 permanent sample plots (PSPs) across the PEF, shown as point symbols in Figure~\ref{pefMap}. The underlying surface in Figure~\ref{pefMap} was generated by passing the point-referenced AGB through a deterministic surface interpolator. Due to MU specific harvesting and subsequent regrowth cycles, the surface exhibits patterns of spatial dependence with relatively strong homogeneity within MUs. For example, MU U7B---highlighted in Figure~\ref{pefMap}---received a shelterwood harvest in 1978 with a final overstory harvest in 2003. This silvicultural treatment results in a MU with relatively young trees and even-aged composition with low AGB (indicated by a lighter surface color in Figure~\ref{pefMap}). In contrast to U7B, C12 is characterized by older and larger trees, but also greater vertical and horizontal forest structure complexity due to repeated selection harvests that aim to concentrate growth on economically desirable trees. \citet{sendak2003} and \citet{hayashi2014} provide additional silvicultural treatment details.

Large footprint waveforms, characteristic of space-based LiDAR sensors, were calculated using discrete multistop returns from a 2013 PEF G-LiHT data acquisition campaign \citep{cook2013}. As noted in Section~\ref{Sec: Intro}, G-LiHT is a portable multi-sensor airborne system developed by NASA Goddard Space Flight Center that simultaneously maps the composition and structure of terrestrial ecosystems. The G-LiHT laser scanner (VQ-480, Riegl Laser Measurement Systems, Horn, Austria) uses a 1550 nm laser that provides an effective measurement rate of up to 150 kHz along a 60$^\circ$ swath perpendicular to the flight direction. At a nominal flying altitude of 335 m, each laser pulse has a footprint approximately 10 cm in diameter and is capable of producing up to 8 returns. Following data processing methods in \cite{blair1999}, G-LiHT produced 26,286 georeferenced pseudo-waveform LiDAR signals across the PEF with 451 of these spatially coinciding with the observed PSPs. Each pseudo-waveform covers a 15 m diameter footprint with a signal comprising 113 values between 0 and 33.9 m above the ground. A signal value is the amount of energy returned to the sensor from a given height divided by the total energy emitted by the sensor over the footprint (additional details are given in Section~\ref{dataPrep}). The signal can be used to characterizes the vertical distribution of forest structure within the footprint. Signals corresponding to PSPs within the MUs highlighted in Figure~\ref{pefMap} are shown in Figure~\ref{pefLidar}. Here, U7B's even-aged and structurally homogeneous composition is apparent in the signals' consistent peak at $\sim$8 m---corresponding to the densest layer of the forest canopy---and minimal energy returns above $\sim$17 m---corresponding to maximum forest canopy height. In contrast, C12's signals are characterized by non-zero values at greater heights---reflecting the prevalence of taller trees---and greater vertical distribution of energy returns---indicative of a vertically complex forest structure resulting from the MU's silvicultural treatments. The relative energy distribution in the signal does not exactly portray the vertical distribution of vegetation because dense overstory may act to reduce the amount of energy available to characterize lower canopy structures. Therefore, if inferential interest is in the vertical distribution of leaf area density, then we would want to transform the signal energy returns to account for decreasing transmittance of energy through the canopy, see, e.g., \citet{macarthur69} for theoretical motivation for such transformations and \citet{stark15} for a recent application. Our focus is on modeling the observed signal and gleaning information from signal characteristics to explain variability in AGB. It is not clear that applying a MacArthur-Horn transformation \citep{macarthur69} to the signal data would fetch improved inference about AGB, and hence we do not pursue such methods here. We do, however, identify these topics as potential extensions to our proposed modeling framework, see Section~\ref{Sec: Summary}.

\section{Models}\label{Sec: Models}
We envision AGB as a continuous spatial process $\{y(\bs) : \bs \in \calD \subset \Re^2\}$ measured over a finite collection of PSP's $\calS = \{\bs_1,\bs_2,\ldots,\bs_{n_s}\} \subset \calD$, where $\calD$ is the domain of interest. LiDAR signals are also assumed to arise as the partial realizations of a process $\{z(\ell) : \ell \in \calD \times \calH\}$, where $\ell = (\bs,x)$ is a space-height coordinate, $z(\ell)$ represents the LiDAR signal's relative energy return at spatial location $\bs$ and height $x$, and $\calH$ is the compact interval $[0,M]$ representing the range of possible heights. The LiDAR signals are also measured at the PSP's in $\calS$ and heights $\calX = \{x_1,x_2,\ldots,x_{n_x}\} \subset \calH$. We will assume that $\calL = \{\ell_1, \ell_2,\ldots,\ell_n\}$ is a complete enumeration of space-height coordinates at which the LiDAR signals have been measured. Each $\ell_i$ will correspond to a unique ordered pair $(\bs_j,x_k)$, where $\bs_j\in\calS$ and $x_k\in\calX$. If the measurements are balanced across space and height, i.e., every PSP has measured the LiDAR signal at each of the points in $\calX$, then there will be $n=n_sn_x$ measurements. This, however, need not be assumed for the subsequent development.  

\subsection{Model for $z(\ell)$}
We write the LiDAR signal as
\begin{align}\label{signal Z}
z(\ell)=\mu_z(\ell;\bbeta_z) + u(\ell) + \epsilon_z(\ell)\;,
\end{align}
where $\mu_z(\ell; \bbeta_z)$ is a mean function capturing large-scale variation, $u(\ell)$ is an underlying zero-centered stochastic process over $\calD\times \calH$ that characterizes spatial dependence, and $\epsilon_z(\ell)\stackrel{ind}{\sim} N(0,\tau_z^2(x))$ models random disturbances at finer scales, at least part of which is attributed to measurement error. The variance of this fine scale disturbance is assumed to remain invariant over the locations, but depends upon the height $x$ at which the signal is measured.  

We assume that $u(\ell)$ is a zero-centered Gaussian process over $\calD\times \calH$ with a covariance function $C_u(\ell, \ell';\boldsymbol{\theta}_u):= \text{Cov}[u(\ell),u(\ell')]$. This function must ensure that the resulting variance-covariance matrix corresponding to realizations of the process over \emph{any} finite subset of $\calD \times \calH$ is positive definite. A natural class of such functions is that of spatiotemporal covariance functions, but with the temporal domain being replaced by the ``height'' domain; \cite{ggg07} and \cite{gnei10} provide excellent expositions of such functions. 

A relevant concern in our current application is the lack of \emph{separability}, i.e., the covariance function should not factorize into a purely spatial component and a purely height component. Separability would imply that the spatial association in the LiDAR signals remains invariant across heights and, similarly, the association among signals at different heights remains the same for each spatial location. This assumption is too stringent for our application; see, e.g., the disparity in empirical semivariogram parameter estimates presented in the Supplementary Material. Furthermore, separable covariance functions violate the so-called ``screening'' effect \citep[][]{Stein_2005} and the resulting associations can be sensitive to small perturbations in spatial locations.    

Based upon the above, we use a slightly simpler version of a highly flexible class of covariance functions developed by \cite{Gneiting_2002},
\begin{align}\label{space-time-cor}
&C_u(\ell,\ell';\boldsymbol{\theta}_u) :=\frac{\sigma_u^2}{(a|x-x'|^{2}+1)^{\gamma}}\exp\bigg(-\frac{c\parallel \mathbf{s}-\mathbf{s}'\parallel}{(a|x-x'|^{2}+1)^{\gamma/2}}\bigg)\;,
\end{align} 
where $\ell = (\bs,x)$, $\ell'= (\bs',x')$,  and $\boldsymbol{\theta}_u=\{\sigma_u^2, a,\gamma,c\}$, with $\sigma_u^2$, $a$, and $c$ all greater than 0 and $\gamma \in [0,1]$. The parameter $\gamma$ describes the space-height interaction. Values of $\gamma$ close to 1 indicate strong space and height interaction. If $\gamma$ is zero, then (\ref{space-time-cor}) is reduced to a separable covariance function, i.e., no space and height interaction. Observe that the above covariance function still assumes isotropy, i.e., the associations depend upon the distances between the spatial locations and the absolute difference between the heights. This, too, is unlikely in practice, but we are less concerned here because nonstationarity will be introduced in the covariance structures as a part of dimension reduction (Section~\ref{Sec: Dim_Red}).   

\subsection{Model for $y(\bs)$ and $z(\ell)$}
\noindent The spatial process for AGB, $y(\bs)$, shares the same spatial domain as the LiDAR process and can be modeled using a Gaussian process over $\calD$. Thus, 
\begin{align}\label{signal Y}
 y(\bs) = \mu_y(\bs;\bbeta_y) + w(\bs) + \epsilon_y(\bs)\; ,
\end{align}
where $\mu_y(\bs;\bbeta_y)$ captures large scale variation or trends in AGB, $w(\bs)$ is a zero-centered spatial process, and $\epsilon_y(\bs)$ is a white noise process with zero mean and variance $\tau_y^2$ to capture measurement error in AGB.

We posit that the process for AGB is associated with the process for the LiDAR signals and desire to estimate this association. 
One possibility is to treat $w(\bs)$ as a shared process between AGB and LiDAR and introduce it as an additive component in (\ref{signal Z}). This, however, causes identifiability issues. First, an additional additive process in (\ref{signal Z}) may be difficult to identify from $u(\ell)$ using a single partial realization of the LiDAR process. Second, the AGB process is then governed by a single shared process, $w(\bs)$, and adding a second process, say $v(\bs)$, to capture departure from the shared component will, again, introduce identifiability problems. Both these problems can, in principle, be resolved in Bayesian settings if prior elicitation was possible on these different component processes. This, unfortunately, is difficult here and we do not pursue such approaches. 

We prefer to treat $u(\ell)$ as a shared underlying process, common to both $z(\bs;x)$ and $y(\bs)$. However, since the AGB has support over the spatial domain only, we assume that it is a continuous weighted average of $u(\ell)$ over $\calX$. Therefore, we write $w(\bs)$ in (\ref{signal Y}) as
\begin{align}\label{Eq: w_s}
 w(\bs) = \int_{\calX} \alpha(x)u(\bs,x)dx + v(\bs) \approx \sum_{j=1}^{n_x} \alpha(x_j)u(\bs,x_j) + v(\bs)\; ,
\end{align}
where $\alpha(x)$ is a weight function that maps height in $\calX$ to the real line and $v(\bs)$ is a zero-centered spatial process, independent of $u(\bs,x)$, that captures features specific to AGB that are not shared with the LiDAR signal. Specifically, we assume $v(\bs)$ is a zero-mean Gaussian process with an exponential covariance function $C_v(\bs,\bs';\btheta_v) = \sigma_v^2 \exp(-\phi_v\|\bs-\bs'\|)$, where $\btheta_v = \{\sigma^2_v,\phi_v\}$. More generally, a Mat\'{e}rn covariance function \citep{Stein_1999} with a prior on the smoothness parameter could have been used, but this does adds to the computational burden without any discernible benefits in the substantive scientific inference we seek in the current application. 

Rather than specify the weights $\alpha(x)$, we represent the integrated process in (\ref{Eq: w_s}) as a linear combination of the $u(\ell)$'s over $\calX$ for any fixed $\bs$ and regard the $\alpha(x_j)$'s as unknown coefficients for the $u(\bs,x_j)$'s. These coefficients capture the dependence of $w(\bs)$ on $u(\ell)$ and, hence, the association between the two processes. If they are all estimated to be effectively zero, then there is no association between the AGB and LiDAR processes, while significant departures of any of the coefficients from zero will indicate association between the processes.

Let $\bu$ be the $n\times 1$ vector with elements $u(\ell_i)$, $i=1,2,\ldots,n$ stacked so that $\ell_i = (\bs_j,x_k)$, where $i = (j-1)n_x + k$ with $j=1,2,\ldots,n_s$ and $k \in \{1,2,\ldots,n_x\}$, and $\bC_u(\btheta_u)$ is the corresponding  $n\times n$ variance-covariance matrix with entries $\mbox{cov}(u(\ell_i), u(\ell_{i'}))$. For the spatial process $v(\bs)$, we let $\bv$ be the $n_s\times 1$ vector with elements $v(\bs_j)$ and $\bC_v(\btheta_v)$ is the corresponding $n_s\times n_s$ spatial covariance matrix. Also, we assume linear fixed effects $\mu_z(\ell_i;\bbeta_z) = \bq_z(\ell_i)^{\top}\bbeta_z$ and $\mu_y(\bs_j;\bbeta_y) = \bq_y(\bs_j)^{\top}\bbeta_y$, where $\bq_z(\ell_i)$ and $\bq_y(\bs_j)$ are $p_z\times 1$ and $p_y\times 1$ vectors of predictors or explanatory variables for $z(\ell_i)$ and $y(\bs_j)$, respectively. 

A joint Bayesian hierarchical model for $y(\bs_j)$'s and $z(\ell_i)$'s, given measurements over $\calS$ and $\calS\times \calX$, respectively, is given by
\begin{align}\label{Eq: Full_BHM}
& p(\bTheta) \times N(\bbeta_y\given \bmu_{\beta_y},\bV_{\beta_y})\times N(\bbeta_z\given \bmu_{\beta_z},\bV_{\beta_z}) \times N(\balpha\given \bmu_{\alpha},\bV_{\alpha}) \times N(\bv\given \bzero, \bC_v(\btheta_v)) \nonumber\\
&\quad \times N(\bu\given \bzero, \bC_u(\btheta_u)) \times \prod_{j=1}^{n_s} N(y(\bs_j)\given \bq_y^{\top}(\bs_j)\bbeta_y + \balpha^{\top}\bu(\bs_j) + v(\bs_j),\tau_y^2) \nonumber\\ 
&\qquad\times \prod_{i=1}^{n} N(z(\ell_i)\given \bq_z^{\top}(\ell_i)\bbeta_z + u(\ell_i),\tau_z^2(x_k)) \; ,
\end{align}
where $\bTheta = \{\btheta_u,\btheta_v, \tau^2_y, \btau^2_z\}$ with $\btau^2_z = (\tau^2_z(x_k))_{k=1}^{n_x}$, $\bbeta_z$ and $\bbeta_y$ are regression slopes for each $\bq_z(\ell_i)$ and $\bq_y(\bs_j)$, respectively, $\bu(\bs_j)$ is the vector with elements $u(\bs_j,x_k)$ for $x_k$'s in $\calX$ yielding LiDAR signals corresponding to $\bs_j$, $\balpha$ is an $n_x\times 1$ vector of unknown coefficients, viz. the $\alpha(x_j)$'s, for the elements in $\bu(\bs_j)$, and $p(\bTheta)$ are joint prior distributions on the process parameters for $u(\bs,x)$ and $v(\bs)$. Further specifications customarily assume that
\begin{align}\label{Eq: Prior_Process_Parameters}
p(\bTheta) \propto p(\btheta_u)\times p(\btheta_v) \times IG(\tau^2_y\given a_{\tau_y}, b_{\tau_y}) \times \prod_{k=1}^{n_x}IG(\tau^2_z(x_k)\given a_{\tau_z}, b_{\tau_z})\;, 
\end{align}
where $p(\btheta_u) = p(a,\gamma,c)\times IG(\sigma_u^2|a_u,b_u)$ and $p(\btheta_v) = p(\phi_v)\times IG(\sigma_v^2|a_v,b_v)$, with $IG$ denoting the inverse-Gamma distribution. When the number of space-height coordinates $n$ is large, estimating (\ref{Eq: Full_BHM}) is computationally expensive and, depending upon the available computational resources, possibly unfeasible. 


\subsection{Predictive process counterparts for dimension reduction}\label{Sec: Dim_Red}
To implement the computations necessary for estimating (\ref{Eq: Full_BHM}) when $n$ is large, we exploit reduced rank processes to achieve dimension reduction. Such processes usually arise as basis expansions of the original process with fewer number of basis functions than the number of data points. This yields ``low-rank'' processes. Every choice of basis functions yields a process and there are far too many choices to enumerate here; see, e.g., \cite{wikle_2011} for an excellent overview of these methods. Here, we opt for a particularly convenient choice, the predictive process \citep[][]{Banerjee_etall_2008, Finley_etall_2009}, which derives the basis functions from taking the conditional expectation of the original process, often called the ``parent'' process, given its realizations over a fixed set of points, often referred to as ``knots.'' 

Let $\calS^*_u = \{\mathbf{s}_{u,1}^*,\mathbf{s}_{u,2}^*,\ldots,\mathbf{s}_{u,n_u^*}^*\}$  and $\calS^*_v=\{\mathbf{s}_{v,1}^*,\mathbf{s}_{v,2}^*,\ldots,\mathbf{s}_{v,n^*_v}^*\}$ be two sets of spatial knots to be used for constructing the predictive process counterparts of $u(\ell)$ and $v(\bs)$, labeled $\tildeu(\ell)$ and $\tildev(\bs)$, respectively. Let $\calX^* = \{x_1^*,x_2^*,\ldots,x_{n_x^*}^*\}$ be a set of knots for heights in the LiDAR signal. Dimension reduction is achieved because the number of knots, i.e., $n_u^*$, $n^*_v$, and $n_x^*$, is much smaller than the original number of observations $n_s$ and $n_x$. Implementation details for the predictive process version of (\ref{Eq: Full_BHM}) used in the subsequent analyses is detailed in the Supplemental Material.

\subsection{Bayesian prediction}\label{sec: Bayesian implementation}
As noted in Section~\ref{Sec: Intro}, we seek predictive inference for $z(\ell_0)$ at any arbitrary space-height coordinate $\ell_0$ and for $y(\bs_0)$ at any arbitrary spatial location $\bs_0$. The posterior predictive distributions and corresponding sampling algorithms that yield this inference are defined in the Supplemental Material. In Section~\ref{Sec: illlustrations}, we use posterior predictive inferences at $i$) unobserved locations to create prediction maps of the LiDAR signals and AGB, and to assess models' predictive performance using holdout set validation, and $ii$) observed locations to provide \emph{replicated} data \citep[see, e.g.,][]{gelman2013} used to assess candidate models' fit.


\section{Data analysis}\label{Sec: illlustrations}
The proposed Markov chain Monte Carlo (MCMC) sampler and prediction algorithms, detailed in the Supplementary Material, were implemented in C++. All code needed to fit the proposed models and reproduce the subsequent results are available in the Supplemental Material. Posterior inference for subsequent analysis were based upon three chains of 50000 MCMC iterations (with a burn-in of 5000 iterations). The computations were conducted on a Linux workstation using two Intel Nehalem quad-Xeon processors. 

In the subsequent simulation experiment Section~\ref{Sec: synthetic} and PEF data analysis Section~\ref{Sec: real} candidate models were compared based on parameter estimates, fit to the observed data, out-of-sample prediction, and posterior predictive distribution coverage. Model choice was assessed using the deviance information criterion or DIC and model complexity $p_D$ \citep{spieg02} and a posterior predictive loss criterion D=G+P \citep{gelf98}, where smaller values of DIC and D indicate preferred models. For both analyses, a 25\% holdout set, comprising locations selected at random, served to assess out-of-sample prediction. Prediction accuracy for the holdout locations was measured using root mean squared prediction error (RMSPE) \citep{rmspe02} as well as CRPS and GRS given in Equation~21~and~27, respectively, in \cite{gneiting2007}. Smaller values RMSPE and CRPS, and larger values of GRS, indicate improved predictive ability. The percent of holdout locations that fell within their respective posterior predictive distribution 95\% CI was also computed along with the average interval width.  

\subsection{Simulation experiment}\label{Sec: synthetic}
\begin{figure}[!htbp]
\centering
\subfigure[]{\includegraphics[width=7cm]{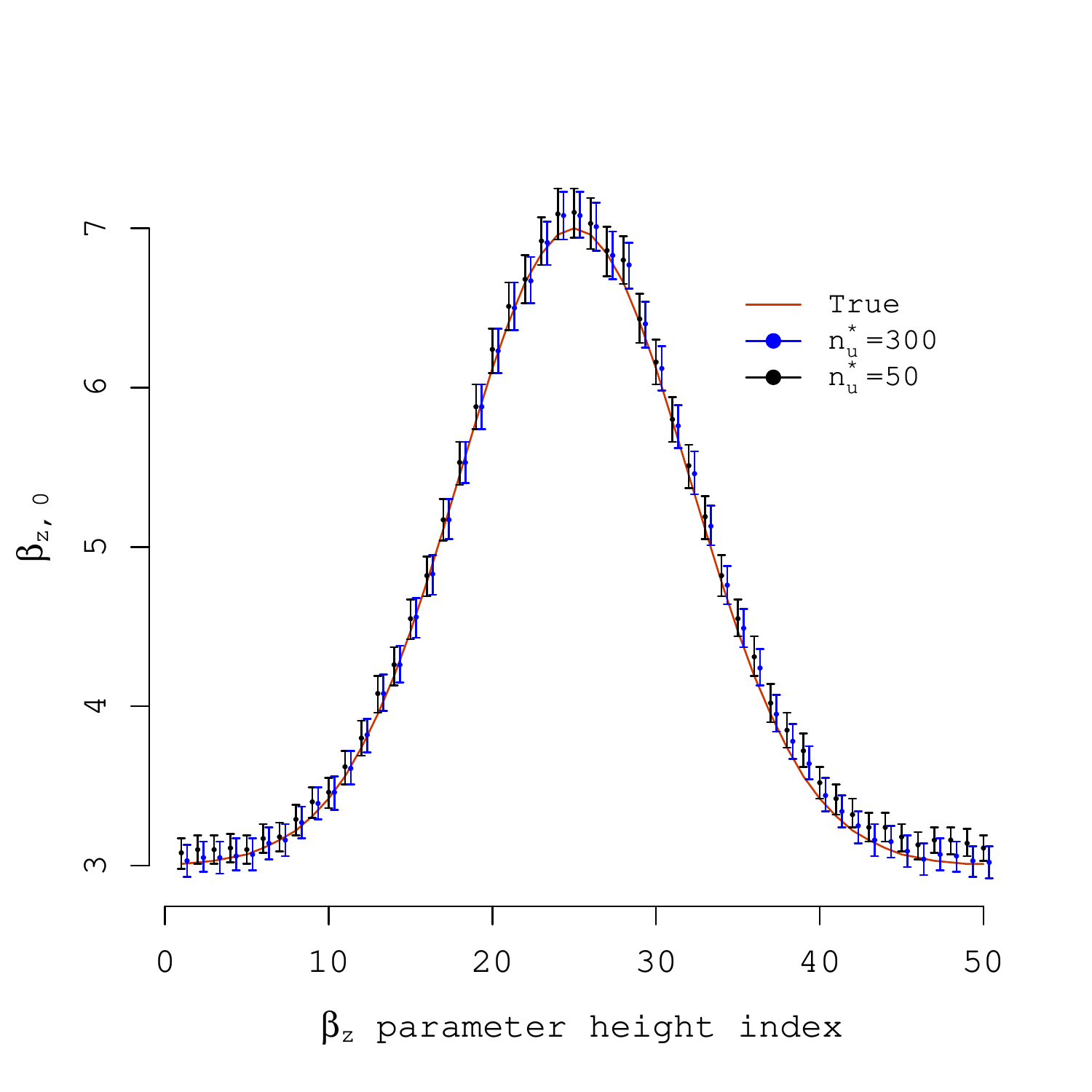}\label{syn-betaZ-est}}
\subfigure[]{\includegraphics[width=7cm]{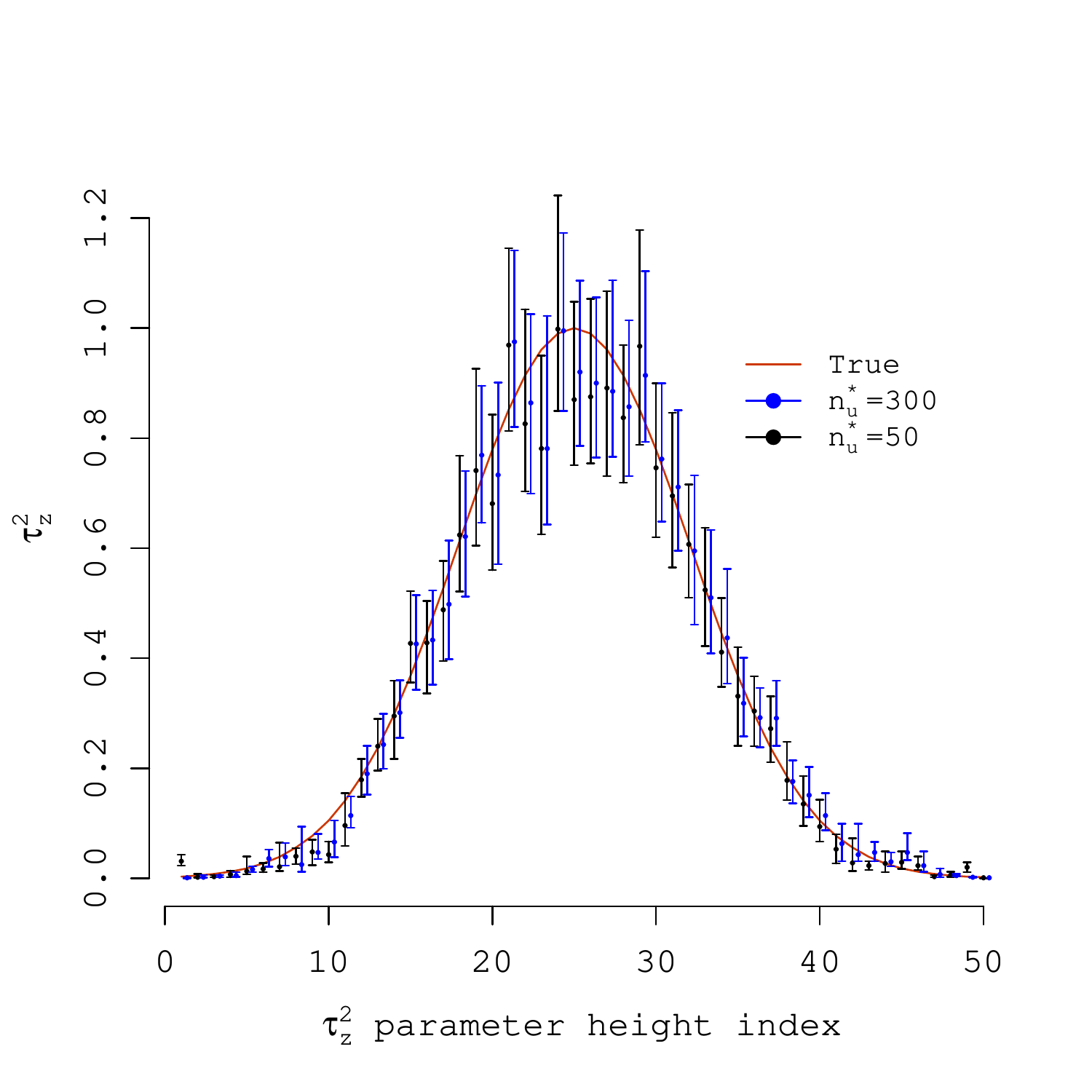}\label{syn-tayZ-est}}
\caption{Parameter posterior summaries, $50\%$ point symbol and $95\%$ credible interval bars. Posterior summaries are jittered slightly along the x-axis to facilitate comparison.}\label{syn-z-ests}
\end{figure}

Using the \emph{true} parameter values given in the first column of Table~\ref{est-pred-simulatedData} and Figure~\ref{syn-z-ests}, we simulated AGB and LiDAR signals from the full GP joint likelihood for AGB and LiDAR in (\ref{Eq: Full_BHM}) for $n_s=400$ coinciding locations in $\calS$ on a regular grid within a $[0,4]\times[0,4]$ domain and $n_x=50$ heights within $[0,5]$. The AGB signal was regressed on a global intercept $(\beta_y)$ while the LiDAR signals were regressed on the $50$ height-specific intercepts and non-spatial variances; thus $\bbeta_z$ and $\btau^2_z$ are both $50\times 1$. A subset of 100 locations from the 400 were withheld to assess out-of-sample predictions. Each of our candidate predictive process models used $n_x^*=5$ equally spaced knots for height in the $[0,5]$ interval and $n_v^*=n_s=300$ with $\calS_v^* = \calS$. Candidate models differed on the number of knots $n_u^*$. We considered models with $n_u^*=300$ and $\calS_u^*=\calS$ and with $n_u^*$ equaling 200, 100, and 50 knots, respectively, selected on a regular grid within the domain. 

\begin{sidewaystable}[!htbp]
\centering
\caption {Parameter credible intervals, $50\%\, (2.5\%,97.5\%)$, goodness-of-fit, and predictive validation. Bold entries indicate where the true value is missed, bold goodness-of-fit metrics indicate \emph{best} fit, and bold prediction metrics indicate \emph{best} predictive performance.}\label{est-pred-simulatedData}
\scriptsize
\begin{tabular}{cccccc}
\toprule
{Parameter} & & \multicolumn{4}{c}{Models} \\
\cmidrule{3-6}
{} & {True} & {$n_u^*=300$ and $\calS_u^*=\calS$} & {$n_u^*=200$} & {$n_u^*=100$}&{$n_u^*=50$} \\
\midrule
$\beta_{y,0}$&20&20.11(19.3,21.06)&20.21(19.48,21.05)&20.28(19.46,21.32)&20.5(19.95,21.12)\\
$\alpha_1$&-2&-1.92(-2.25,-1.61)&-2.09(-2.49,-1.73)&-2.03(-2.63,-1.45)&-1.41(-2.19,-0.71)\\
$\alpha_2$&0&0.06(-0.3,0.46)&0.38(-0.13,0.83)&0.25(-0.62,1.03)&-0.1(-1.09,0.87)\\
$\alpha_3$&2&1.57(1.06,2.01)&\textbf{1.21(0.71,1.77)}&\textbf{1.12(0.26,1.99)}&0.91(-0.09,2.01)\\
$\alpha_4$&1&0.91(0.54,1.29)&0.77(0.34,1.19)&0.93(0.26,1.6)&0.68(-0.19,1.59)\\
$\alpha_5$&5&4.94(4.64,5.23)&4.9(4.52,5.27)&4.82(4.23,5.43)&5.04(4.38,5.68)\\
$\sigma^2_u$&0.2&0.2(0.19,0.22)&0.21(0.2,0.22)&\textbf{0.22(0.21,0.23)}&0.21(0.2,0.22)\\
$a$&12&12.91(11.62,14.46)&11.53(10.1,13.36)&10.58(9.29,12.63)&10.62(8.77,13.19)\\
$\gamma$&0.9&0.89(0.84,0.93)&0.94(0.85,0.97)&0.92(0.85,0.95)&0.92(0.83,0.97)\\
$c$&5&5.17(4.62,5.93)&5.24(4.7,5.87)&\textbf{3.82(3.35,4.29)}&\textbf{3.09(2.7,3.5)}\\
$\sigma^2_v$&0.5&0.64(0.34,1.27)&0.68(0.37,1.24)&0.51(0.15,1.47)&0.6(0.12,1.74)\\
$\phi_v$&2&1.68(0.41,3.87)&1.7(0.49,3.8)&1.38(0.21,5.26)&7.6(0.25,9.99)\\
\midrule
$p_D$ for AGB \& LiDAR &&95.01&95.94&96.14&92.39\\
DIC for AGB \& LiDAR&&\textbf{28399.73}&29238.34&30140.16&30647.04\\
G for AGB \& LiDAR&&6883.54&7705.30&8566.07&8779.24\\
P for AGB \& LiDAR&&7507.49&8215.54&9007.16&9386.19\\
D=G+P for AGB \& LiDAR&&\textbf{14391.03}&15920.84&17573.23&18165.43\\
\midrule
RMSPE for AGB \& LiDAR &&\textbf{0.80}&0.80&0.81&0.82\\
CRPS for AGB \& LiDAR &&\textbf{2054.52}&2056.32&2076.22&2090.47\\
GRS for AGB \& LiDAR &&\textbf{-712.04}&-743.83&-864.92&-962.09\\
95\% prediction interval coverage for AGB\& LiDAR  \%&&95.27&95.1&95.39&94.9\\
\midrule
RMSPE for $\mbox{AGB}\given \mbox{observed LiDAR}$ &&2.79&\textbf{2.74}&2.78&2.81\\
CRPS for $\mbox{AGB}\given \mbox{observed LiDAR}$ &&159.32&\textbf{156.28}&157.71&158.69\\
GRS for $\mbox{AGB}\given \mbox{observed LiDAR}$ &&-311.06&\textbf{-305.18}&-308.63&-308.17\\
95\% prediction interval coverage for $\mbox{AGB}\given \mbox{observed LiDAR}$ $\%$&&89&91&93&93\\
95\% prediction interval width for $\mbox{AGB}\given \mbox{observed LiDAR}$&&9.30&9.70&9.74&9.96\\
CPU time &&101.7 h&60.0 h &27.5 h&13.8 h\\
\bottomrule
\end{tabular} 
\end{sidewaystable}

Parameter estimates and performance metrics for all candidate models are given in Table~\ref{est-pred-simulatedData}. With the exception of $\alpha_3$ for $n_u^*$ equal to 200 and 100, and a few of the covariance parameters for $n_u^*$ equal to 100 and 50, the 95\% CIs for all parameters included the \emph{true} values. Importantly, the $\alpha$ estimates---used to relate information between LiDAR signals and AGB---remain consistent in sign and magnitude as the spatial process associated with the signals is modeled over a reduced number of knots. Figure~\ref{syn-z-ests} provides the posterior summaries for the $50$ height-specific intercepts and non-spatial variances associated with the LiDAR signals; results for only two candidate models are provided due to the large number of parameters and minimal difference in estimates among the models. These estimates also seem robust to a coarser representation of the underlying process (Figure~\ref{syn-z-ests}). 

Not surprisingly, for the joint outcome vector, goodness-of-fit and out-of-sample prediction is best for the full model, i.e., $n_u^*=300$ (rows labeled  AGB \& LiDAR in Table~\ref{est-pred-simulatedData}). Interestingly, in an interpolation setting when LiDAR is observed, RMSPE, CRPS and GRS all show that AGB prediction improves slightly when moving from the full model to the $n_u^*=200$ knot model (rows labeled  AGB $|$ observed LiDAR in Table~\ref{est-pred-simulatedData}). In general, goodness-of-fit and predictive performance is not substantially degraded for the predictive process models when compared to the full model. The last row in Table~\ref{est-pred-simulatedData} gives the CPU time for the candidate models. A 6-fold decrease in knots between the full model and $n_u^*=50$ knot model results in a 7-fold decrease in computing time.

\subsection{Forest LiDAR and biomass data analysis}\label{Sec: real}

\subsubsection{Data preparation and exploratory data analysis}\label{dataPrep}
Pre-processing the raw G-LiHT LiDAR data followed methods detailed in \cite{cook2013} and produced a complete $15\times 15$ m grid across the PEF, where each pixel contained a LiDAR signal. Prior to analysis, these LiDAR signals were further processed to remove excess zeros and coarsened to remove small-scale noise. Specifically, the maximum tree height across the PEF was approximately 22.8 m and hence LiDAR signal values beyond this height were zero and subsequently removed. Small-scale anomalies that occurred across each signal's 113 values were smoothed by averaging every two consecutive measurements. Truncation above forest canopy extent and smoothing resulted in signals of length $n_x=39$ within the $[0, 22.8]$ m height interval. Figure~\ref{pefLidar} illustrates the processed signals over the PSPs within two MUs. 

As described in Section~\ref{Sec: Intro}, important current and future LiDAR acquisition missions sparsely sample the domain of interest. The sampling designs, e.g., transects or clusters, aim to collect LiDAR data across the domain and also at forest inventory plot locations. To mimic the sparseness of these anticipated datasets and associated inferential challenges, only LiDAR signals that spatially coincided with PSPs were used for candidate model parameter estimation. 

Candidate models were assessed without and with predictor variables. The set of candidate models without predictor variables (i.e., $\bq_z=\bones$ and $\bq_y=1$, i.e., intercepts only) mimic a worst-case settings where we do not have complete coverage, wall-to-wall, predictor variables. The set of candidate models with predictor variables use ground surface topographic characteristics derived from G-LiHT's digital elevation model to help explain variability in AGB and LiDAR signals. Although we considered a host of aspect, slope, and roughness predictor variables in exploratory data analysis using the proposed models \citep[following suggested topographic transformations in][]{stage76}, only elevation consistently explained a substantial portion of variability in observed AGB and LiDAR signals. Therefore, the set of models with predictor variables was fit using $\bq_z^{\top}(\ell)=(1, \text{Elev}(\bs))$ and $\bq_y^{\top}(\bs)=(1, \text{Elev}(\bs))$, where Elev is ground elevation (m).

To better assess the information contribution of latent LiDAR regressors for AGB prediction an additional set of intercept only models were fit with $v(\bs)$ set to zero. For this set of models, only information from the latent LiDAR regressors, i.e., via $\balpha$, is available to explain variability in AGB.

\begin{figure}[!ht]
\begin{center}
\includegraphics[width=10cm]{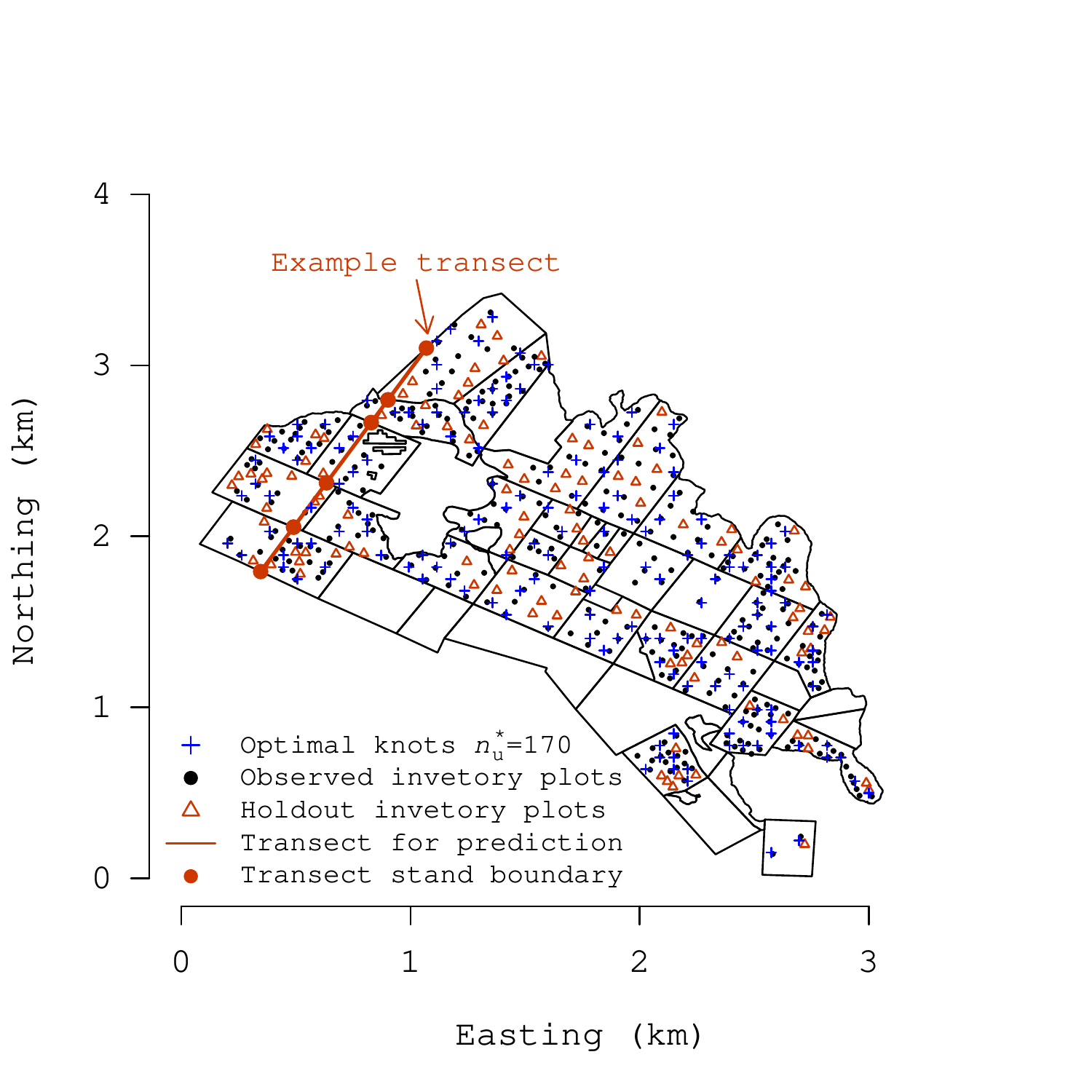}
\caption{Locations of observed and holdout PSPs, predictive process knots, and illustrative transect on the PEF.}\label{PEF-transects-holdout}
\end{center}
\end{figure}

\subsubsection{Candidate model results and discussion}
For brevity, in the main text we only present results for the set of candidate models what include elevation as a predictor variable. Results for the intercept only and $v(\bs)$ set to zero models are offered in the Supplemental Material. Results were comparable among all candidate model sets; however, models with the elevation predictor and $v(\bs)$ showed consistent, albeit marginal, improvement in fit and predictive performance.

Candidate models were formed by varying $n^*_x$, $n^*_u$, and $n^*_v$ along with knot location following the approximately optimal knot design criteria described in the Supplementary Material. We present results for $n^*_x$ from 2 to 7, $n^*_u$ equal 339 and 170, and $n^*_v=339$. All candidate models were fit using a subset of $n_s=339$ PSPs selected at random from the complete set of 451 PSPs. The remaining 112 PSP were used for out-of-sample prediction validation. Observed and holdout PSPs along with knot locations are illustrated in Figure~\ref{PEF-transects-holdout}. Here too, an example transect is identified along with locations where the transect crosses MU boundaries. This example transect is used to help visualize and assess results.

Parameter estimates and goodness-of-fit metrics for the aforementioned choices of $n^*_x$ are provided in Tables~\ref{est-pred-realData-339-with-v} and \ref{est-pred-realData-170-with-v} for $n^*_u$ equal 339 and 170, respectively. For both choices of $n^*_u$, increasing the number of height knots $n^*_x$ resulted in improved fit (noted by lower values of DIC and D). This makes sense because a greater number of knots provides an improved representation of the LiDAR signal. This result also holds for the intercept only and $v(\bs)=0$ candidate models (Supplementary Material Tables~\ref{est-pred-realData-339-with-v}, \ref{est-pred-realData-170-with-v}, \ref{est-pred-realData-339-wo-v}, and \ref{est-pred-realData-170-wo-v}). 

\begin{figure}[!htbp]
\centering
\subfigure[]{\includegraphics[width=5.25cm]{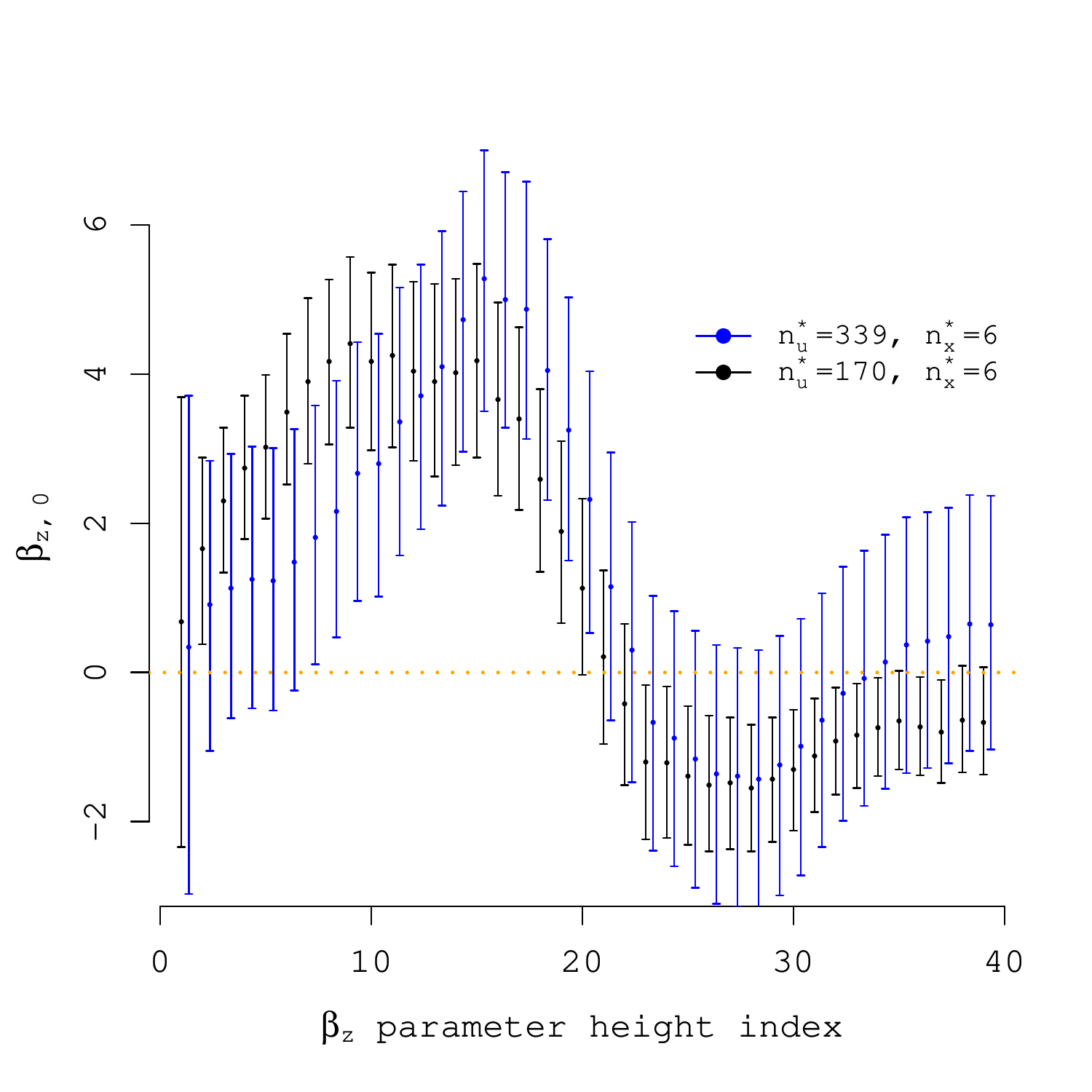}\label{pef-beta0Z-est}}
\subfigure[]{\includegraphics[width=5.25cm]{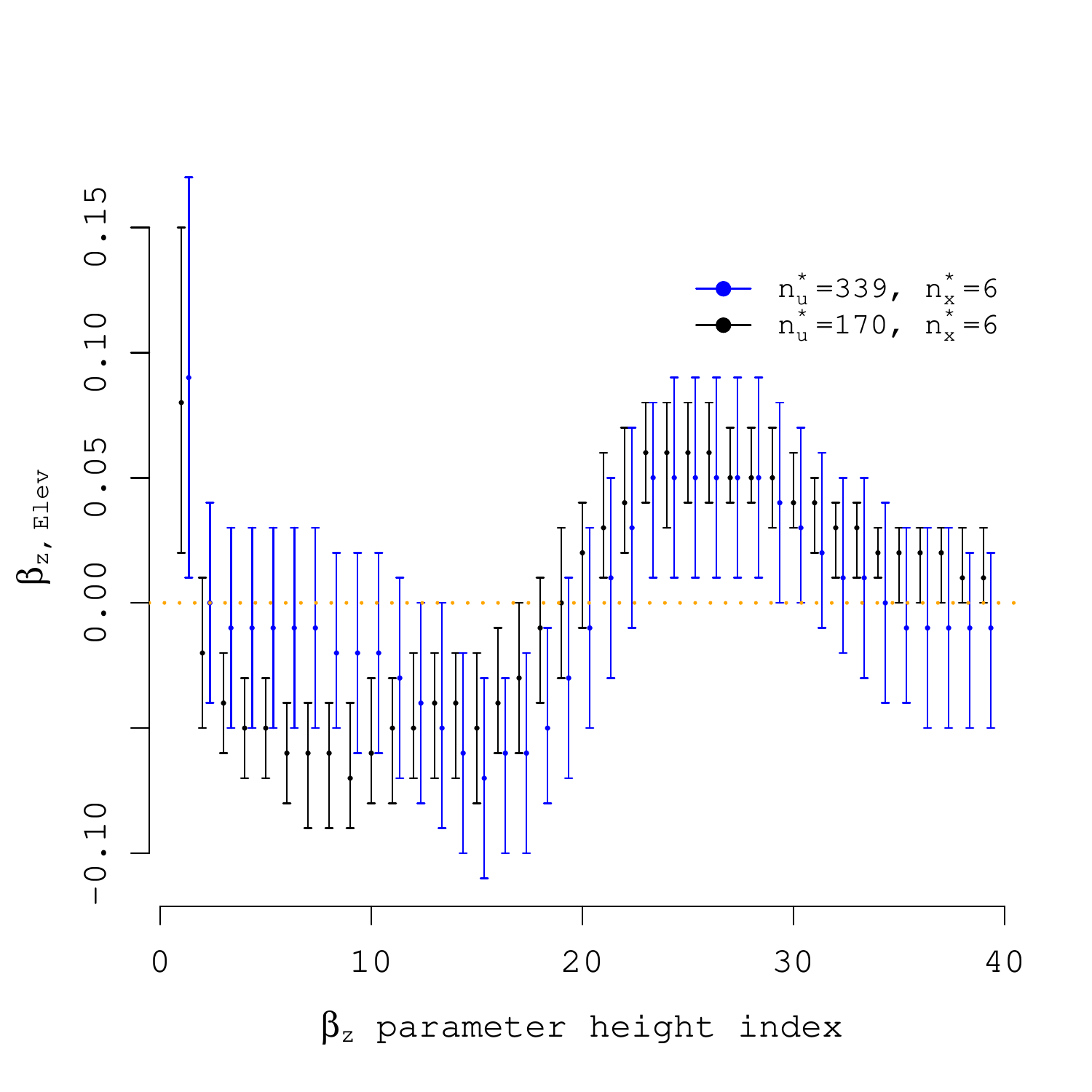}\label{pef-beta1Z-est}}
\subfigure[]{\includegraphics[width=5.25cm]{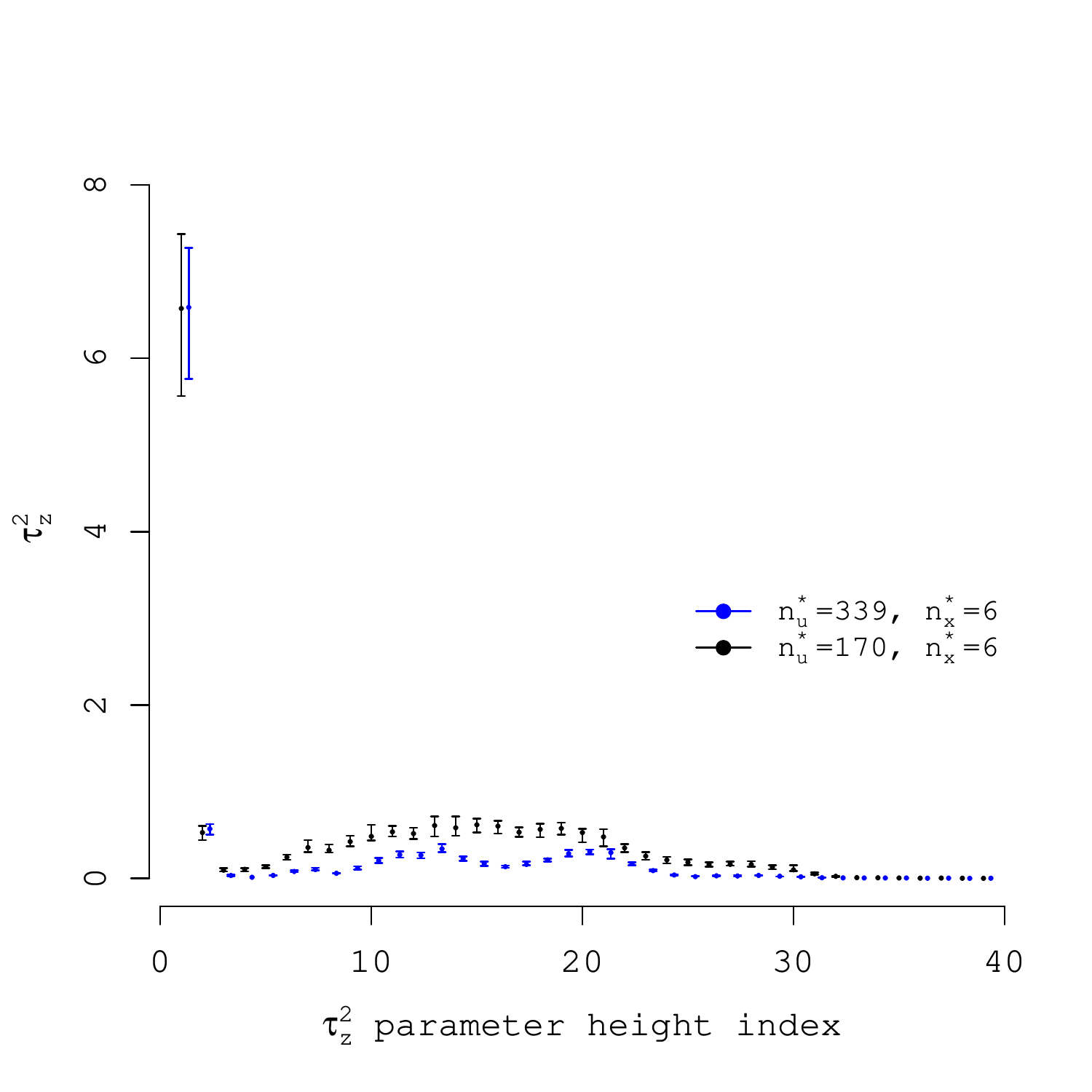}\label{pef-tayZ-est}}
\caption{Parameter posterior summaries, $50\%$ point symbol and $95\%$ credible interval bars. Posterior summaries are jittered slightly along the x-axis to facilitate comparison.}\label{pef-z-ests}
\end{figure}

The regression slope parameter estimates for $\beta_{y,\text{Elev}}$ in Tables~\ref{est-pred-realData-339-with-v} and \ref{est-pred-realData-170-with-v} suggest elevation explains a significant amount of variation in AGB, with greater biomass occurring on higher elevation PSPs. Figure~\ref{pef-z-ests} provides posterior summaries for the LiDAR signal's height-specific intercept $\beta_{z,0}$, elevation regression slope parameter $\beta_{z,\text{Elev}}$, and non-spatial variance parameter estimates for two candidate models. For heights of less than $\sim$10 m (corresponding to $\bbeta_z$ index $\sim$20 in Figure~\ref{pef-beta1Z-est}), higher elevation is associated with fewer energy returns, and between $\sim$10-18 m (corresponding to $\bbeta_z$ index $\sim$20-35 in Figure~\ref{pef-beta1Z-est}) higher elevation is associated with greater energy returns. This is not surprising, given $\beta_{y,\text{Elev}}$ estimates suggest greater AGB is associated with higher elevation and tall dense canopies are indicative of forest with greater AGB.

Also, unlike the synthetic data analysis, Figures~\ref{pef-beta0Z-est} and \ref{pef-beta1Z-est}, show differences in precision between parameter estimates at different levels of $n^*_u$. Specifically, we see more precise estimates of the intercept and elevation regression slope parameters at $n^*_u=170$ versus $n^*_u=339$. This is likely due to a phenomena called spatial confounded \citep[see, e.g.,][]{hanks15}, which is most pronounced when a predictor variable and random effect are correlated, in our case elevation seems to be correlated spatially with $\tildeu(\ell)$. The greater the resolution on the spatial process, i.e. moving from $n^*_u=170$ to 339, the greater the influence of spatial confounding on the estimates of $\beta_{1,\text{Elev}}$. Spatial confounding can result in wider regression coefficient credible intervals but should not have deleterious effects on prediction, i.e., the inferential focus of our analysis. 

Figure~\ref{pef-tayZ-est} shows lower residual variances for $n^*_u=339$ across heights. This is not surprising, given the additional information about the signal supplied by the higher resolution spatial process representation. 

\begin{sidewaystable}[!htbp]
\caption {Parameter credible intervals, $50\%\, (2.5\%,97.5\%)$, and goodness-of-fit for the $n_u^*=339$ and $n_v^*=339$ models. Bold parameter values indicated values that differ from zero where appropriate and bold goodness-of-fit metrics indicate \emph{best} fit.}\label{est-pred-realData-339-with-v}
\scriptsize
\begin{tabular}{ccccccc}
\toprule
{Parameter}  & \multicolumn{6}{c}{Height knot models} \\
\cmidrule{2-7}
{}  & {$n_x^*=2$} & {$n_x^*=3$} & {$n_x^*=4$}&{$n_x^*=5$}&{$n_x^*=6$}&{$n_x^*=7$} \\
\midrule
$\beta_{y,0}$&0.21(-0.43,0.86)&0.01(-0.7,0.76)&-0.06(-0.77,0.65)&0.39(-0.51,1.35)&0.09(-0.76,0.98)&-0.12(-1.01,0.71)\\
$\beta_{y,\text{Elev}}$&\textbf{0.02(0,0.03)}&\textbf{0.02(0.01,0.04)}&\textbf{0.03(0.01,0.04)}&0.02(-0.01,0.04)&\textbf{0.02(0,0.04)}&\textbf{0.03(0.01,0.05)}\\
$\alpha_1$&\textbf{-0.08(-0.15,-0.01)}&-0.06(-0.15,0.03)&-0.04(-0.09,0.01)&0.02(-0.04,0.07)&-0.05(-0.1,0.01)&-0.04(-0.09,0.02)\\
$\alpha_2$&\textbf{0.33(0.27,0.39)}&0.06(-0.11,0.24)&\textbf{-0.16(-0.23,-0.1)}&-0.05(-0.11,0.03)&\textbf{-0.11(-0.18,-0.04)}&\textbf{-0.07(-0.13,0)}\\
$\alpha_3$&\qquad -&\textbf{0.43(0.33,0.52)}&\textbf{0.1(0.05,0.16)}&0.05(-0.02,0.12)&\textbf{-0.1(-0.17,-0.03)}&\textbf{-0.09(-0.15,-0.03)}\\
$\alpha_4$&\qquad -&\qquad -&\textbf{0.26(0.13,0.39)}&\textbf{0.2(0.14,0.27)}&0.04(-0.03,0.12)&-0.01(-0.06,0.06)\\
$\alpha_5$&\qquad -&\qquad -&\qquad -&\textbf{0.28(0.13,0.47)}&\textbf{0.11(0.03,0.18)}&0.06(-0.02,0.12)\\
$\alpha_6$&\qquad -&\qquad -&\qquad -&\qquad -&\textbf{0.18(0.06,0.35)}&\textbf{0.17(0.07,0.28)}\\
$\alpha_7$&\qquad -&\qquad -&\qquad -&\qquad -&\qquad -&0.04(-0.13,0.28)\\
$\sigma^2_u$&0.1(0.1,0.11)&0.16(0.14,0.17)&0.48(0.44,0.52)&0.61(0.57,0.65)&0.97(0.88,1.07)&1.15(1.07,1.26)\\$a$&1.12(0.99,1.28)&1.41(1.24,1.56)&0.75(0.7,0.8)&1.12(1.05,1.18)&0.99(0.93,1.05)&1.05(0.99,1.11)\\
$\gamma$&0.99(0.98,0.99)&1(1,1)&1(0.99,1)&1(1,1)&0.99(0.98,1)&1(1,1)\\
$c$&16.39(14.03,19.62)&10.94(9.89,12.14)&8.89(8.16,9.72)&8.55(7.98,9.1)&8.07(7.54,8.59)&8.44(7.94,9.01)\\
$\sigma^2_v$&0.09(0.07,0.15)&0.11(0.07,0.18)&0.1(0.06,0.13)&0.1(0.07,0.17)&0.08(0.07,0.11)&0.1(0.08,0.12)\\
$\phi_v$&3.6(2.23,7.49)&3.27(1.86,4.93)&3.59(1.84,5.32)&2.14(1.64,3.55)&4.51(3.34,6.58)&3.8(3.04,4.82)\\
$\tau_y^2$&0.03(0.01,0.04)&0.03(0.02,0.04)&0.03(0.02,0.03)&0.03(0.02,0.04)&0.02(0.01,0.03)&0.02(0.02,0.03)\\ 
\midrule
$p_D$&112.67&113.14&105.05&104.81&104.28&100.52\\
DIC&25249.95&23914.36&19508.47&15141.14&12678.65&\textbf{9627.62}\\
G&10201.51&8243.9&6932.84&4466.97&3636.59&3165.57\\
P&10606.52&8770.24&7522.97&4918.89&4188.48&3595.77\\
D&20808.03&17014.13&14455.8&9385.86&7825.07&\textbf{6761.33}\\
\bottomrule
\end{tabular}
\end{sidewaystable}

\begin{sidewaystable}[!htbp]
\caption {Parameter credible intervals, $50\%\, (2.5\%,97.5\%)$, and goodness-of-fit for the $n_u^*=170$ and $n_v^*=339$ models. Bold parameter values indicated values that differ from zero where appropriate and bold goodness-of-fit metrics indicate \emph{best} fit.}\label{est-pred-realData-170-with-v}
\scriptsize
\begin{tabular}{ccccccc}
\toprule
{Parameter}  & \multicolumn{6}{c}{Height knot models} \\
\cmidrule{2-7}
{}  & {$n_x^*=2$} & {$n_x^*=3$} & {$n_x^*=4$}&{$n_x^*=5$}&{$n_x^*=6$}&{$n_x^*=7$} \\
\midrule
$\beta_{y,0}$&-0.09(-0.76,0.59)&-0.44(-1.16,0.27)&-0.25(-0.88,0.37)&-0.15(-0.9,0.57)&-0.15(-0.82,0.49)&-0.2(-0.92,0.48)\\
$\beta_{y,\text{Elev}}$&\textbf{0.03(0.01,0.04)}&\textbf{0.03(0.02,0.05)}&\textbf{0.03(0.02,0.04)}&\textbf{0.03(0.01,0.04)}&\textbf{0.03(0.01,0.04)}&\textbf{0.03(0.01,0.04)}\\
$\alpha_1$&-0.2(-0.42,0.04)&\textbf{-0.4(-0.66,-0.14)}&-0.03(-0.17,0.1)&-0.07(-0.22,0.07)&-0.1(-0.26,0.03)&-0.13(-0.32,0.08)\\
$\alpha_2$&\textbf{0.72(0.53,0.89)}&-0.45(-0.79,0.01)&\textbf{-0.28(-0.5,-0.08)}&\textbf{-0.25(-0.44,-0.06)}&\textbf{-0.13(-0.28,0)}&-0.09(-0.26,0.09)\\
$\alpha_3$&\qquad -&\textbf{0.62(0.37,0.86)}&\textbf{0.23(0.05,0.41)}&-0.13(-0.31,0.05)&\textbf{-0.28(-0.44,-0.14)}&\textbf{-0.24(-0.45,-0.04)}\\
$\alpha_4$&\qquad -&\qquad -&\textbf{0.33(0.08,0.55)}&\textbf{0.24(0.11,0.42)}&0.14(-0.03,0.31)&-0.06(-0.2,0.1)\\
$\alpha_5$&\qquad -&\qquad -&\qquad -&0.1(-0.19,0.41)&0.11(-0.05,0.31)&0.12(-0.06,0.32)\\
$\alpha_6$&\qquad -&\qquad -&\qquad -&\qquad -&0.09(-0.22,0.31)&0.12(-0.12,0.37)\\
$\alpha_7$&\qquad -&\qquad -&\qquad -&\qquad -&\qquad -&0.05(-0.3,0.41)\\
$\sigma^2_u$&0.05(0.05,0.06)&0.09(0.07,0.1)&0.24(0.22,0.28)&0.35(0.32,0.39)&0.39(0.35,0.44)&0.42(0.38,0.48)\\
$a$&0.8(0.64,1)&1.02(0.82,1.23)&0.84(0.74,0.94)&1.23(1.12,1.35)&1.25(1.14,1.37)&1.49(1.36,1.64)\\
$\gamma$&0.98(0.9,0.99)&0.99(0.96,1)&0.99(0.97,1)&0.99(0.98,1)&1(0.99,1)&1(0.99,1)\\
$c$&6.02(5.32,6.71)&4.75(4.19,5.39)&3.23(2.85,3.6)&3.05(2.77,3.39)&2.81(2.53,3.15)&2.89(2.55,3.29)\\
$\sigma^2_v$&0.08(0.05,0.16)&0.09(0.05,0.14)&0.06(0.04,0.09)&0.08(0.04,0.18)&0.05(0.04,0.08)&0.06(0.04,0.1)\\
$\phi_v$&2.63(1.32,5.02)&2.51(1.43,4.54)&3.92(2.02,8.4)&1.42(1.13,3.61)&8.17(3.76,16.33)&4.1(2.6,7.84)\\
$\tau_y^2$&0.04(0.02,0.05)&0.03(0.02,0.04)&0.03(0.02,0.04)&0.04(0.02,0.05)&0.02(0.01,0.05)&0.03(0.02,0.04)\\ 
\midrule
$p_D$&118.32&121.42&110.41&112.36&114.27&110.13\\
DIC&26952.72&26222.81&24446.58&22809.18&22325.5&\textbf{21889.75}\\
G&10703.8&9529.07&8085.34&6678.27&6402.89&6131.33\\
P&10868.98&9840.64&8480.8&7080.34&6885.51&6656.74\\
D&21572.78&19369.71&16566.14&13758.61&13288.4&\textbf{12788.07}\\
\bottomrule
\end{tabular}
\end{sidewaystable}

Inference on $\balpha$, which act as weights for the LiDAR process, help us gauge the usefulness of the latent LiDAR regressors for explaining variability in AGB. The increasing subscript value on $\alpha$'s in Tables~\ref{est-pred-realData-339-with-v} and \ref{est-pred-realData-170-with-v} correspond to increasing knot heights in $\calX^*$. Regardless of the choice for $n^*_u$ or $n^*_v$, estimates of $\alpha$ and knot height are positively associated, for example, estimates of $\alpha$ for the $n^*_x=6$ model in Table~\ref{est-pred-realData-339-with-v} increase from $\alpha_1$ (corresponding to knot $x^*_1$ which is near the ground) to $\alpha_6$ (corresponding to $x^*_6$ which is near the maximum forest canopy height). The intuition here is that the latent LiDAR process $\tildeu(\ell)$ tend to have larger values at heights where energy return is greater (i.e., where the signal encounters tree material such as leaves, branches, boles) and small where energy return is low (i.e., where there is mostly empty space in the vertical profile of the forest, or dense overstory intercepts the majorly of the signal). Typically, more mature forest with large diameter and tall trees have higher AGB compared with younger lower canopy or sparsely populated forest. Therefore, we expect greater AGB in regions returning much of the LiDAR signal from greater heights and, conversely, lower AGB in regions returning much of the signal at lower heights.

Tables~\ref{est-pred-realData-339-with-v-2} and \ref{est-pred-realData-170-with-v-2} provide out-of-sample prediction validation results corresponding to the models presented in Tables~\ref{est-pred-realData-339-with-v} and \ref{est-pred-realData-170-with-v}, respectively. For joint prediction of AGB and LiDAR, using either level of $n^*_u$, RMSPE favors $n^*_x=7$ whereas selection results based on CRPS and CRS are mixed. Importantly, however, holdout validation results suggest there is very little difference in predictive ability among the range of height knots beyond $n^*_x=2$ or 3. If interest is in predicting AGB at a location given observed LiDAR, the majority of the prediction metrics favor $n^*_x$ between 4 and 6 as indicated in the lower portion of Tables~\ref{est-pred-realData-339-with-v-2} and \ref{est-pred-realData-170-with-v-2}.

\begin{table}[!htbp]
\begin{center}
\caption {Prediction metrics for the $n_u^*=339$ and $n_v^*=339$ models. Bold values indicate \emph{best} predictive performance.}\label{est-pred-realData-339-with-v-2}
\scriptsize
\begin{tabular}{ccccccc}
\toprule
{Parameter}  & \multicolumn{6}{c}{Height knot models} \\
\cmidrule{2-7}
{}  & {$n_x^*=2$} & {$n_x^*=3$} & {$n_x^*=4$}&{$n_x^*=5$}&{$n_x^*=6$}&{$n_x^*=7$} \\
\midrule
RMSPE for AGB\&$\mbox{LiDAR}$&0.883&0.842&0.83&0.781&0.777&\textbf{0.77}\\
CRPS for AGB\&$\mbox{LiDAR}$&1944.36&1868.3&1846.69&\textbf{1754.68}&1773.21&1777.23\\
GRS for AGB\&$\mbox{LiDAR}$&-1895.37&-1381.9&-1359.45&\textbf{-1084.02}&-1230.43&-1376.24\\
95\% prediction coverage for AGB\&$\mbox{LiDAR}$&93.5&93.8&94.2&93.4&94.9&95.7\\ \midrule
RMSPE for $\mbox{AGB}\given \mbox{observed LiDAR}$ &0.313&0.308&\textbf{0.304}&0.311&0.306&0.305\\
CRPS for $\mbox{AGB}\given \mbox{observed LiDAR}$ &19.84&19.36&\textbf{19.16}&19.49&19.26&19.22\\
GRS for $\mbox{AGB}\given \mbox{observed LiDAR}$ &140&151.73&\textbf{154.43}&151.4&153.75&153.93\\
95\% prediction interval coverage for $\mbox{AGB}\given \mbox{observed LiDAR}$ &90.2&93.8&93.8&96.4&95.5&95.5\\
95\% prediction interval width for $\mbox{AGB}\given \mbox{observed LiDAR}$ &1&1.1&1.12&1.23&1.19&1.22\\
\bottomrule
\end{tabular} 
\end{center}
\end{table}

\begin{table}[!htbp]
\begin{center}
\caption {Prediction metrics for the $n_u^*=170$ and $n_v^*=339$ models. Bold values indicate \emph{best} predictive performance.}\label{est-pred-realData-170-with-v-2}
\scriptsize
\begin{tabular}{ccccccc}
\toprule
{Parameter}  & \multicolumn{6}{c}{Height knot models} \\
\cmidrule{2-7}
{}  & {$n_x^*=2$} & {$n_x^*=3$} & {$n_x^*=4$}&{$n_x^*=5$}&{$n_x^*=6$}&{$n_x^*=7$} \\
\midrule
RMSPE for AGB\&$\mbox{LiDAR}$&0.876&0.846&0.824&0.798&0.793&\textbf{0.792}\\
CRPS for AGB\&$\mbox{LiDAR}$&1916.97&1867.7&1814.08&1755.98&1748.6&\textbf{1741.51}\\
GRS for AGB\&$\mbox{LiDAR}$&-1009.11&-1052.07&\textbf{-976.27}&-1060.06&-1024.2&-1103.49\\
95\% prediction coverage for AGB\&$\mbox{LiDAR}$&94.6&94.2&93&91.8&91.6&91.5\\ \midrule
RMSPE for $\mbox{AGB}\given \mbox{observed LiDAR}$ &0.304&0.302&0.298&0.303&\textbf{0.296}&0.298\\
CRPS for $\mbox{AGB}\given \mbox{observed LiDAR}$ &19.11&18.92&18.74&19.02&\textbf{18.62}&18.75\\
GRS for $\mbox{AGB}\given \mbox{observed LiDAR}$ &153.35&156.75&158.25&154.26&\textbf{160.24}&158.42\\
95\% prediction interval coverage for $\mbox{AGB}\given \mbox{observed LiDAR}$ &92.9&93.8&93.8&92.9&93.8&94.6\\
95\% prediction interval width for $\mbox{AGB}\given \mbox{observed LiDAR}$ &1.05&1.09&1.05&1.06&1.07&1.08\\
\bottomrule
\end{tabular} 
\end{center}
\end{table}

Results for the intercept only and $v(\bs)=0$ candidate models are presented in the Supplementary Material. For the $v(\bs)=0$ models, fit and prediction is only influenced by the choice of $n^*_x$ and $n^*_u$. Comparing results between the intercept only model and $v(\bs)=0$ suggest that inclusion of $v(\bs)$ has little effect on the \emph{best} model selected using goodness-of-fit and out-of-sample prediction validation metric within each model set, i.e., $n^*_u$ equal 339 and 170 (Supplementary Material Tables \ref{est-pred-realData-339-with-v}-\ref{est-pred-realData-170-wo-v-2}). This suggests that the underlying process seems to be driven by features shared between AGB and LiDAR and there is negligible information on features specific to AGB that are not shared by LiDAR. Hence, the shared AGB-LiDAR process pursued in this work. However, inclusion of $v(\bs)$ does marginally improve fit to the data and prediction. This improvement suggests there is some spatial structure in the residuals of AGB that is not captured by information from the LiDAR signals.

\begin{figure}[!htbp]
\centering
\subfigure[Observed LiDAR signal]{\includegraphics[width=12cm]{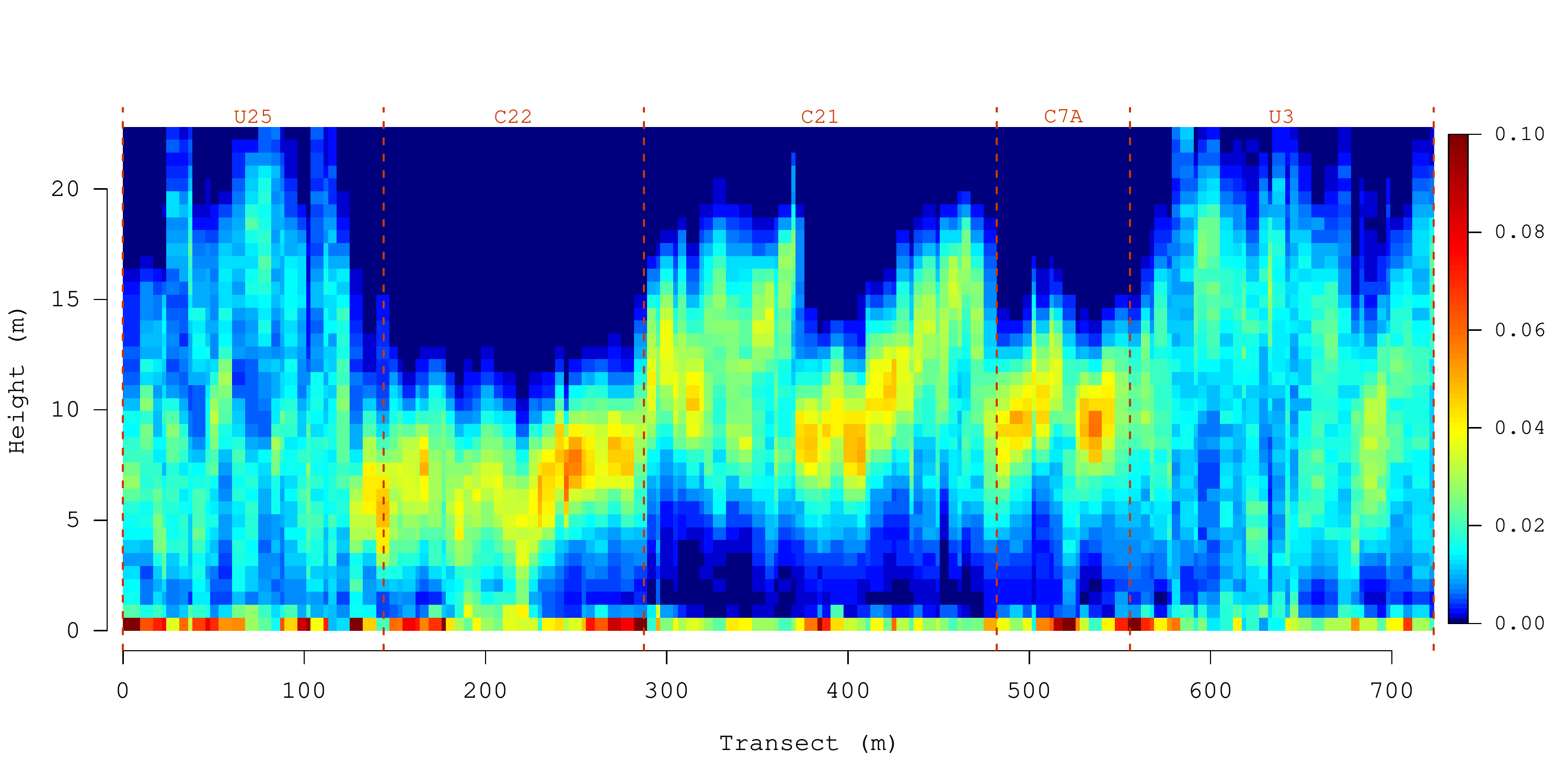}\label{trans-z-obs}}\\
\subfigure[Predicted LiDAR signal with $n_u^*=339$, $n_x^*=6$]{\includegraphics[width=12cm]{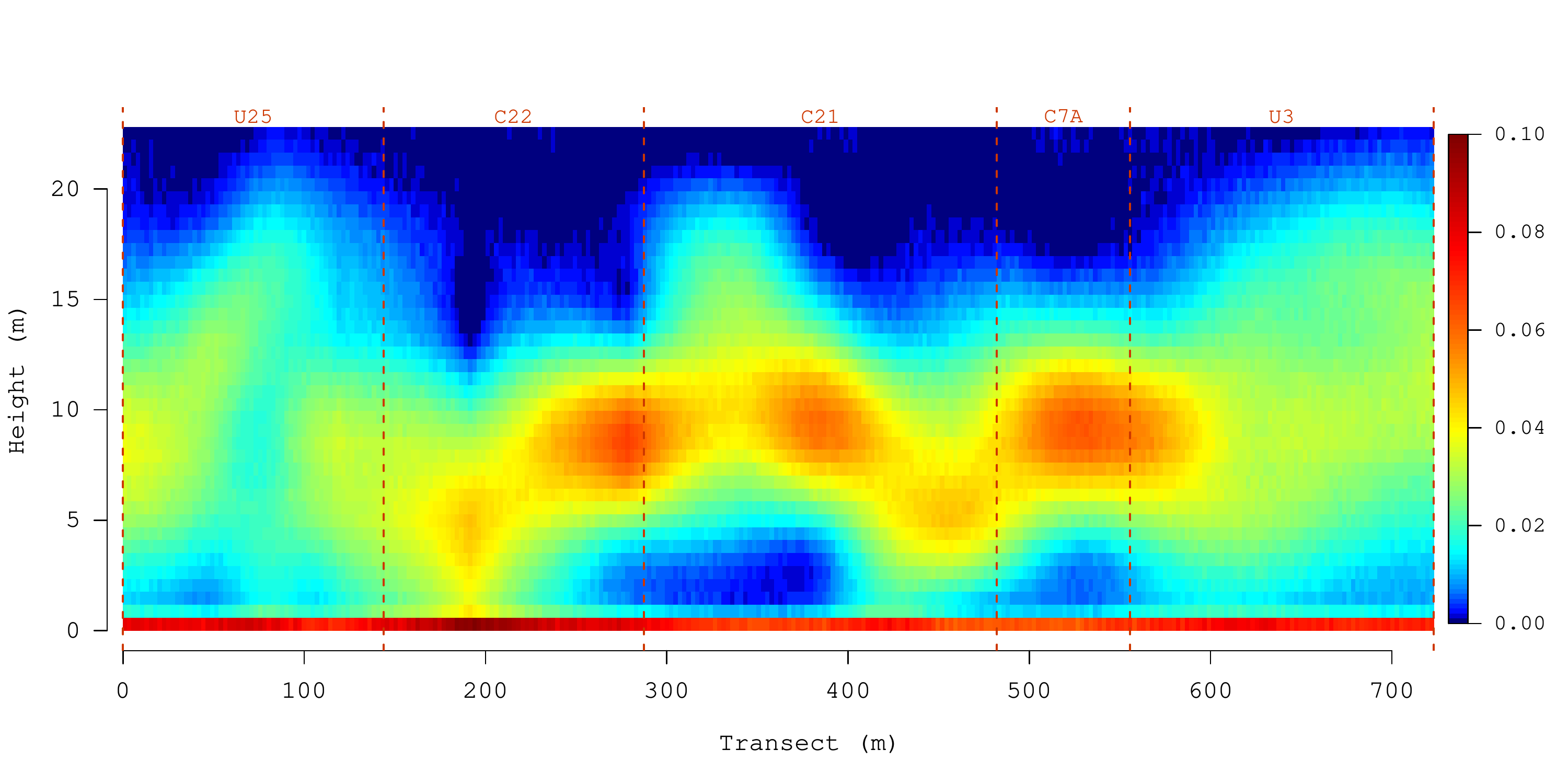}\label{trans-z-339}}\\
\subfigure[Predicted LiDAR signal with $n_u^*=170$, $n_x^*=6$]{\includegraphics[width=12cm]{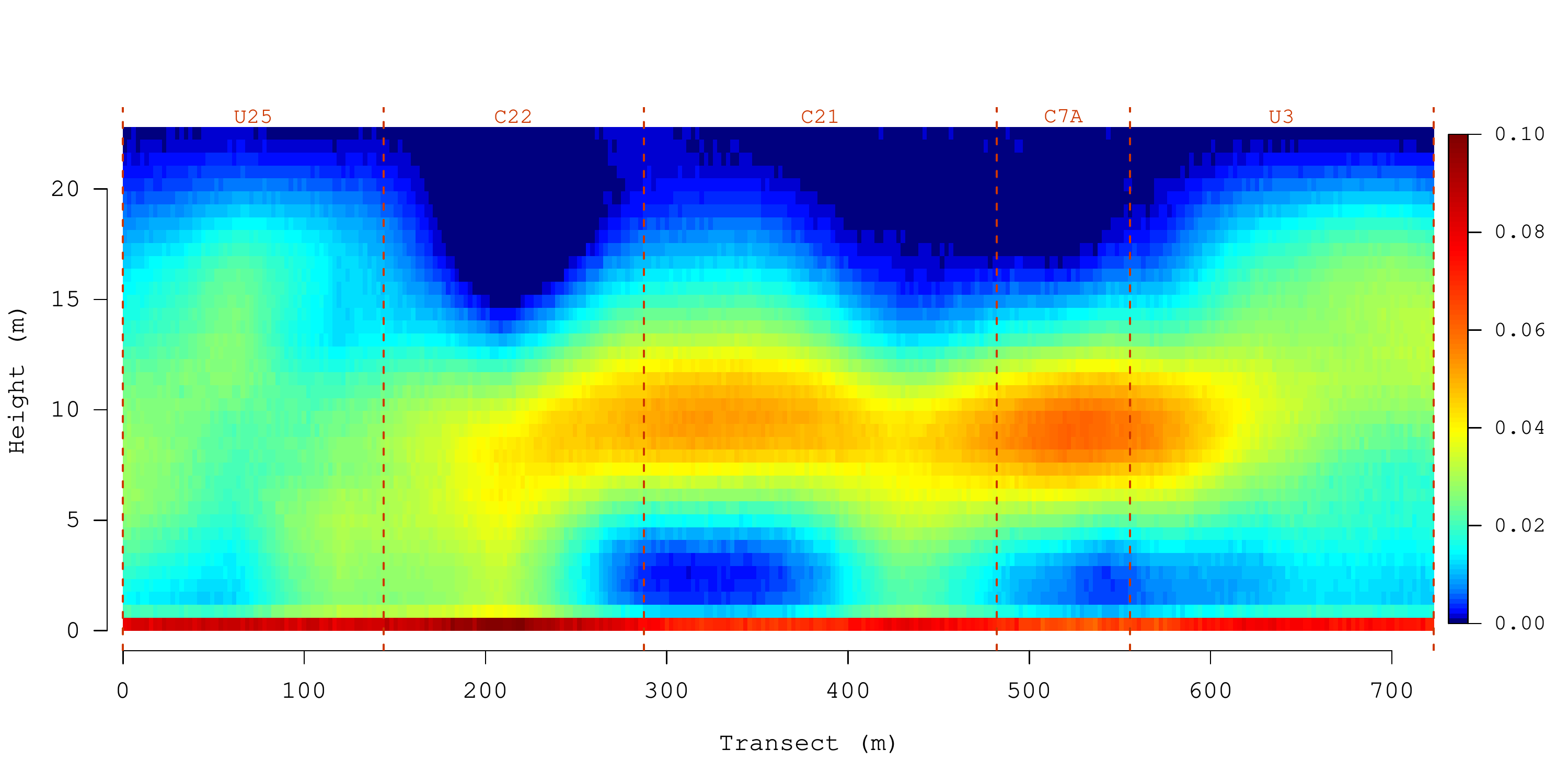}\label{trans-z-170}}\\
\caption{Posterior predictive median for LiDAR signals along the example transect denoted in Figure~\ref{PEF-transects-holdout}.}\label{trans-z}
\end{figure}

It is useful to consider a 2-dimensional slice through the data to further assess candidate model results. Figure~\ref{trans-z-obs} is the side-view of the observed LiDAR signals along the example transect denoted in Figure~\ref{PEF-transects-holdout}. Analogous to the portrayal in Figure~\ref{pefLidar}, larger values of the signal correspond to greater density of tree material; hence, one could imagine Figure~\ref{trans-z-obs} is like looking at the side of a forest (15 m in width and $\sim$700 m in length) where greater values correspond to denser forest. Lower values in Figure~\ref{trans-z-obs} could occur because there is no forest (i.e., above the canopy extent), sparse forest, or overstory acts to block the LiDAR signal from penetrating into the lower portions of the forest. MU boundaries are also superimposed on Figure~\ref{trans-z-obs} and clearly show how different silvicultural treatments (i.e., tree harvesting) result in different vertical and horizontal distribution of tree material. For example, MU C22 is a young, short stature, forest versus the older, taller, and more vertically homogeneous forest in MU U3. As seen in Figure~\ref{PEF-transects-holdout}, no PSPs or knots in $\calS^*_u$ fall on the transect and hence the LiDAR signals in Figure~\ref{trans-z-obs} were not used for parameter estimation. 

For brevity, we subsequently consider $n^*_x=6$ candidate models but note that, in general, values of $n^*_x>=4$ yield comparable results. Signal prediction along the example transect using the $n^*_x=6$ models are given in Figures~\ref{trans-z-339} and \ref{trans-z-170}. Comparison between these predictions and observed data, Figure~\ref{trans-z-obs}, shows the candidate models capture the dominant trends in LiDAR signals. Reducing $\bu^*$ process knots by half, i.e., moving from the $n^*_u=339$ to $n^*_u=170$ model, does not greatly affect the vertical and horizontal distribution of predicted signal values. 

\begin{figure}[!htb]
\begin{center}
\includegraphics[height=8cm,width=8cm]{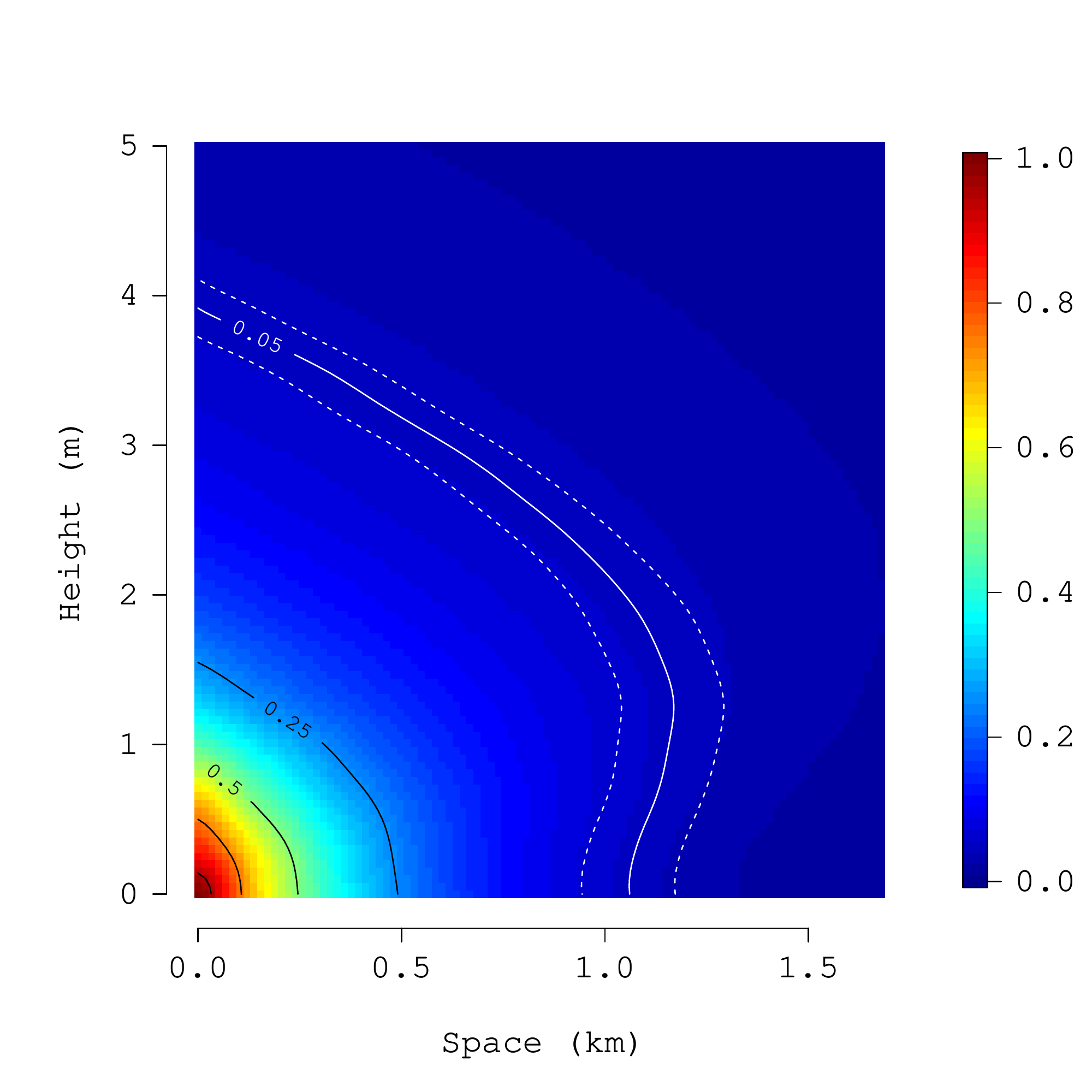}
\caption{Space-height correlation posterior median surface and contours. Median
(solid white lines) and associated 95\% credible interval (dotted white lines) for
 0.05 correlation contour.}\label{cor}
\end{center}
\end{figure}

The observed signal data, Figure~\ref{trans-z}, suggest a strong space-height process interaction. The strength of this interaction is captured by $\gamma$ in covariance function (\ref{space-time-cor}), where values close to one indicate strong interaction and values close to zero indicate weak interaction. Parameter estimates for $\gamma$ in Tables~\ref{est-pred-realData-339-with-v}~and~\ref{est-pred-realData-170-with-v} do indeed corroborate the presence of strong interaction between space and height. Figure~\ref{cor} summarizes estimated space-height correlation and shows the median posterior correlation surface and associated contours using posterior samples from the $n_u^*=170$ and $n^*_x=6$ model. Here, at a given height the spatial correlation is small (i.e., 0.25) at $\sim$0.5 km and negligible (i.e., 0.05) at $\sim$1 km. This makes sense because the average areal extent of the MUs is a bit less than a half kilometer. Within a given signal, i.e., at a given spatial location, the correlation drops to 0.05 at $\sim$4 m. Again, looking at Figures~\ref{pefLidar} and \ref{trans-z-obs}, we see fairly weak correlation in any given signal beyond several meters for most MUs.

\begin{figure}[!htbp]
\centering
\subfigure[Predicted latent LiDAR process with $n_u^*=339$, $n_x^*=6$]{\includegraphics[width=12cm]{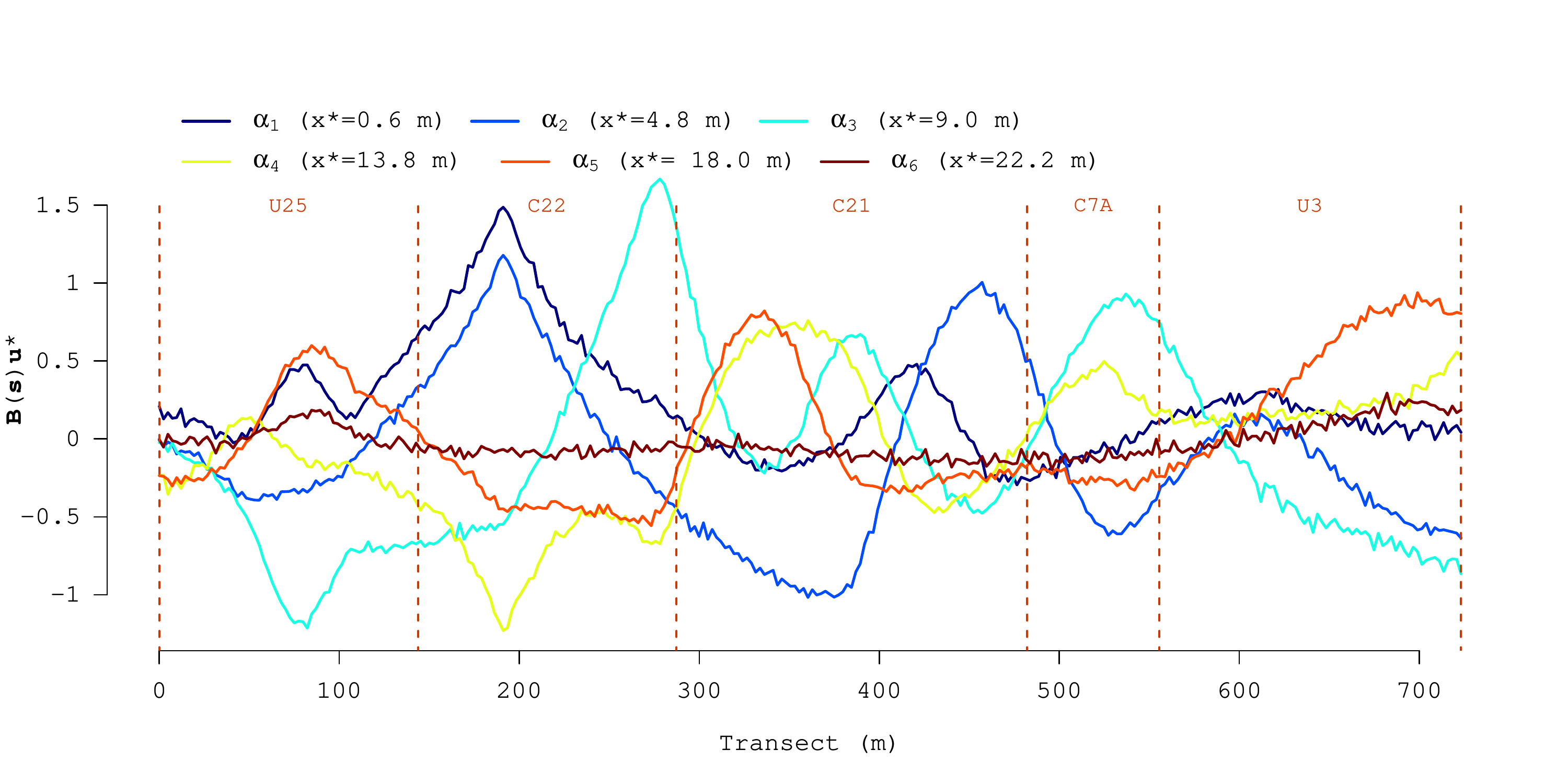}\label{trans-alpha-339}}\\
\subfigure[Predicted latent LiDAR process with $n_u^*=170$, $n_x^*=6$]{\includegraphics[width=12cm]{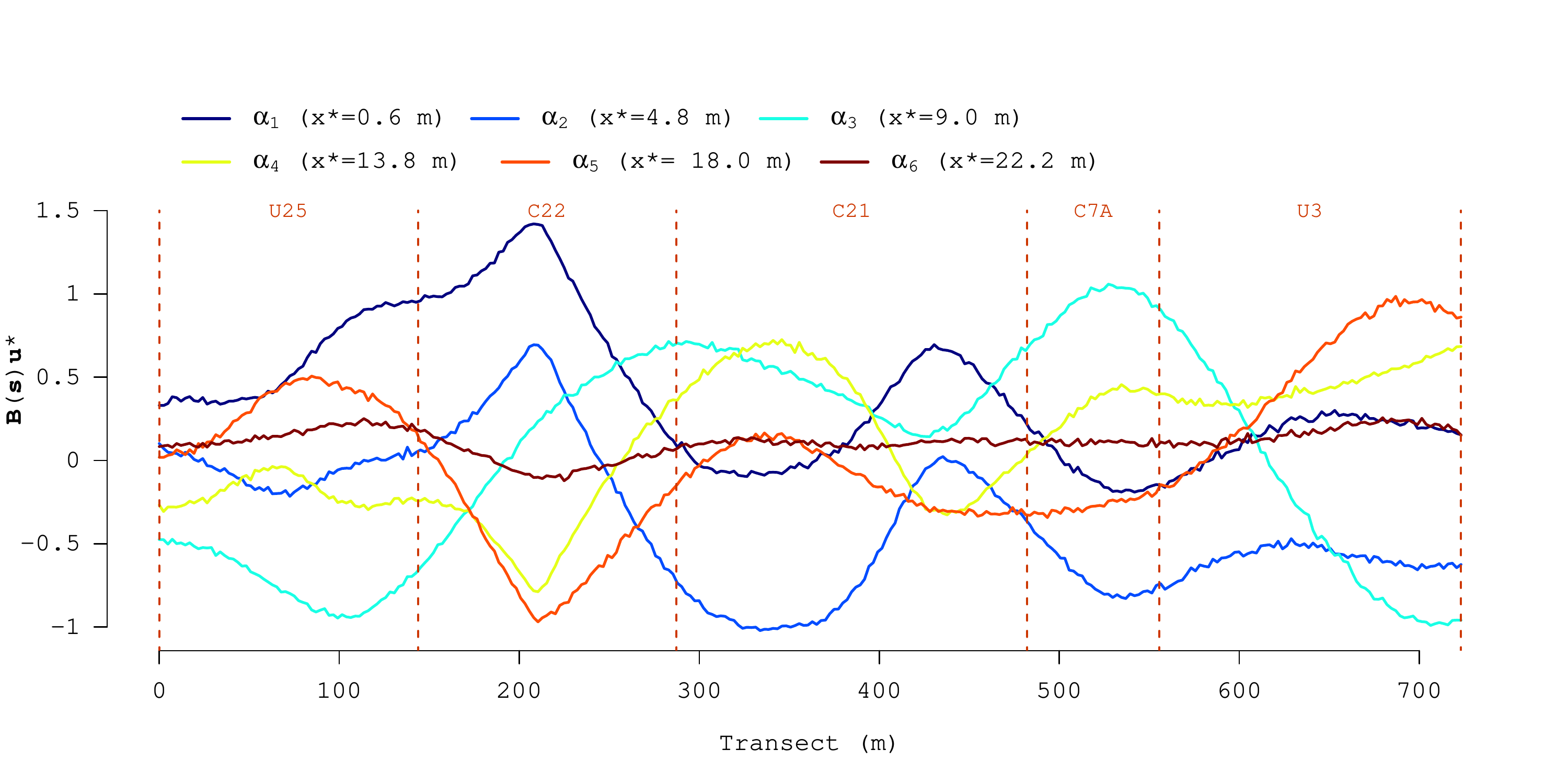}\label{trans-alpha-170}}\\
\caption{Posterior predictive median of latent LiDAR signal regressors along the example transect denoted in Figure~\ref{PEF-transects-holdout}. The legend relates each regressor to the corresponding element in $\balpha$ with the predictive process knot height in meters is given in parentheses.}\label{trans-alpha}
\end{figure}

Figures~\ref{trans-alpha-339} and \ref{trans-alpha-170} display the posterior median for each latent LiDAR regressor along the example transect that correspond to the $n^*_x=6$ model $\alpha$ estimates in Tables~\ref{est-pred-realData-339-with-v}~and~\ref{est-pred-realData-170-with-v}, respectively. Figure legends also include the $x^*$ knot height associated with each latent regressor. The latent regressors interpretation becomes clear when Figure~\ref{trans-z-obs} is considered with Figures~\ref{trans-alpha-339} or \ref{trans-alpha-170}. For example, Figure~\ref{trans-z-obs} shows most energy returns in MU C7A (between $\sim$490-560 m along the example transect) are between $\sim$5-12 m in height, hence we see large values of the latent regressor associated with $x^*=9$ m and $\alpha_3$ in Figures~\ref{trans-alpha-339} and \ref{trans-alpha-170}. Similarly, paucity of energy returns in the $\sim$1-5 m height at $\sim$300-370 m along the example transect, results in small values of the latent regressors associated with $x^*$ equal to 0.6, and 4.8 m. Deviations seen between the latent regressor lines in Figures~\ref{trans-alpha-339} and \ref{trans-alpha-170} and trends in Figure~\ref{trans-z-obs} are due to process smoothing that results from lack of PSPs and $\calS^*_u$ occurring on the example transect. This smoothing also accounts for difference between Figures~\ref{trans-alpha-339} and \ref{trans-alpha-170}.

\begin{figure}[!htbp]
\centering
\subfigure[Predicted AGB with $n_u^*=339$, $n_x^*=6$]{\includegraphics[width=12cm]{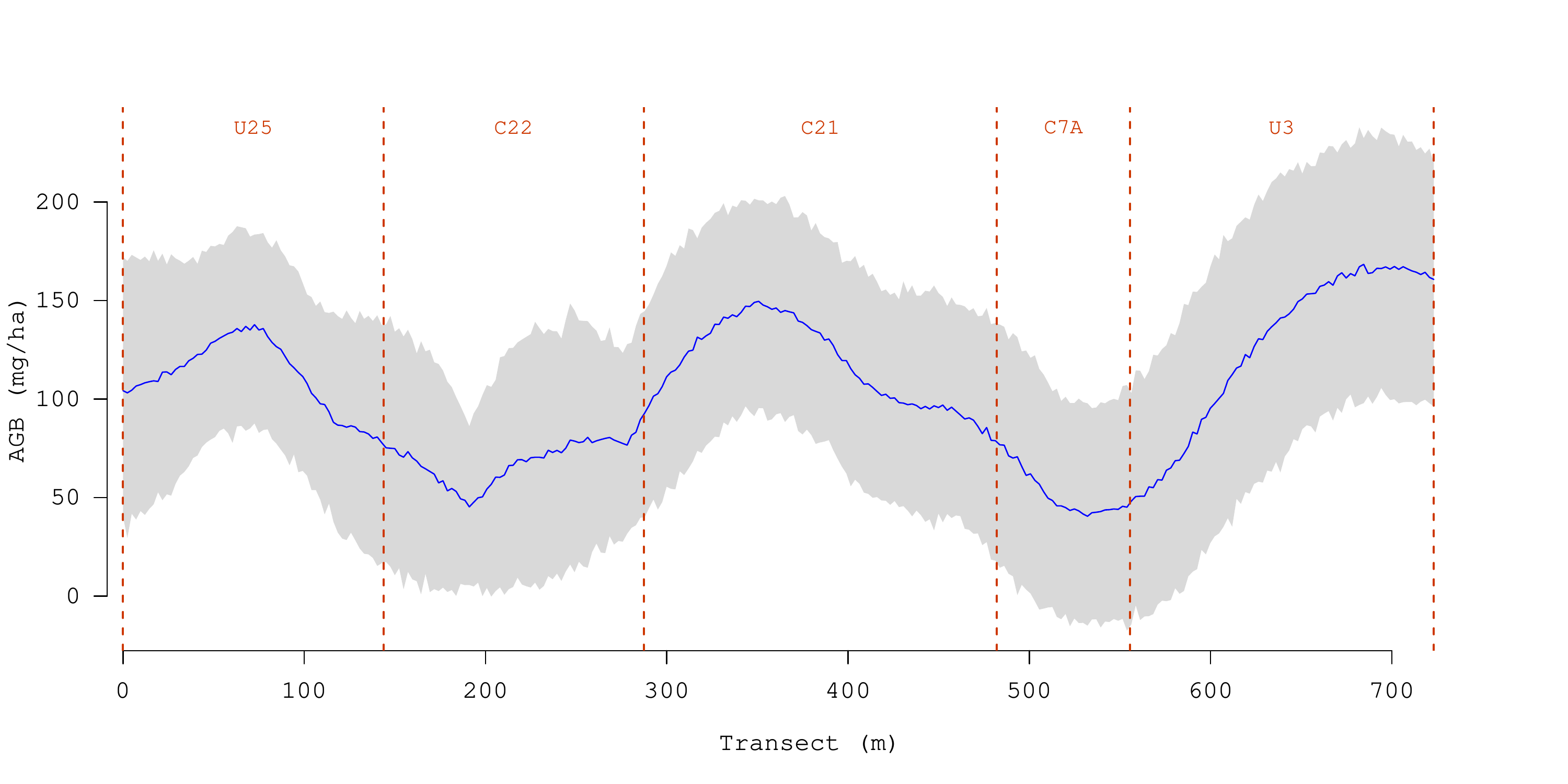}}\\
\subfigure[Predicted AGB with $n_u^*=170$, $n_x^*=6$]{\includegraphics[width=12cm]{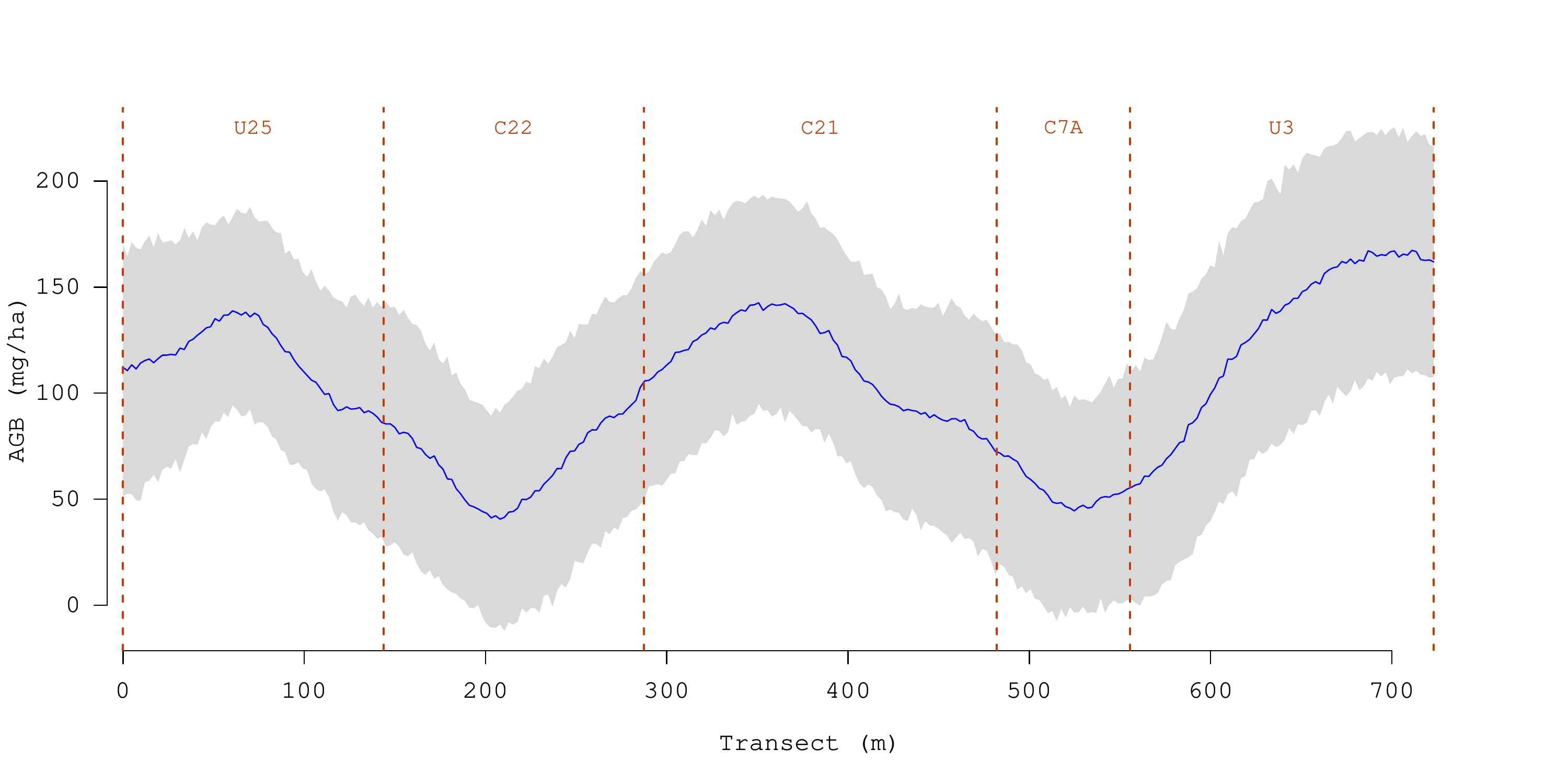}}\\
\caption{Posterior predictive median for AGB along the example transect denoted in Figure~\ref{PEF-transects-holdout}. MU identifiers are provided across the top of each panel.}\label{trans-y}
\end{figure}

\begin{figure}[!htbp]
\centering
\subfigure[Predicted AGB with $n_u^*=339$, $n_x^*=6$]{\includegraphics[width=8cm]{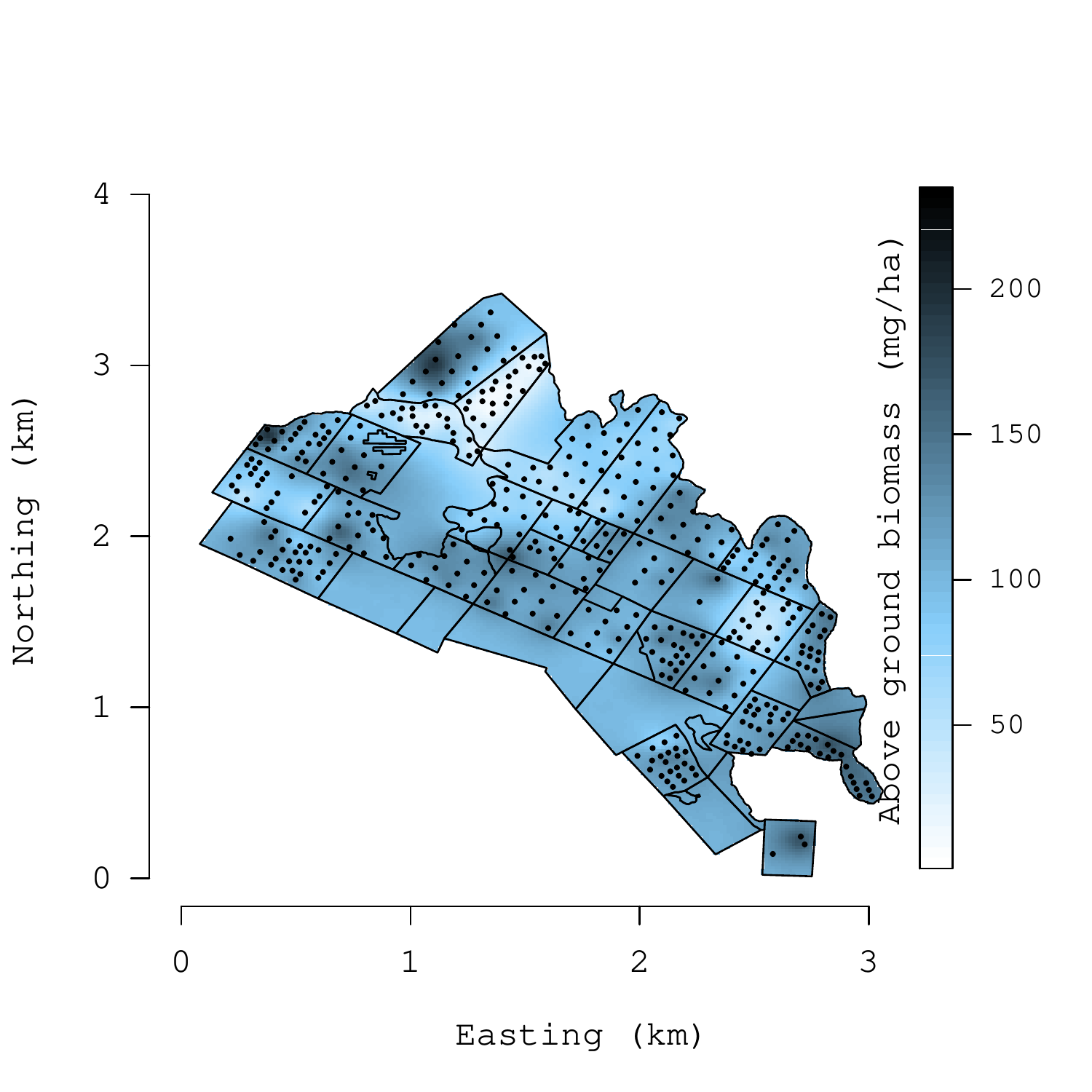}}
\subfigure[95\% prediction interval width for AGB with $n_u^*=339$, $n_x^*=6$]{\includegraphics[width=8cm]{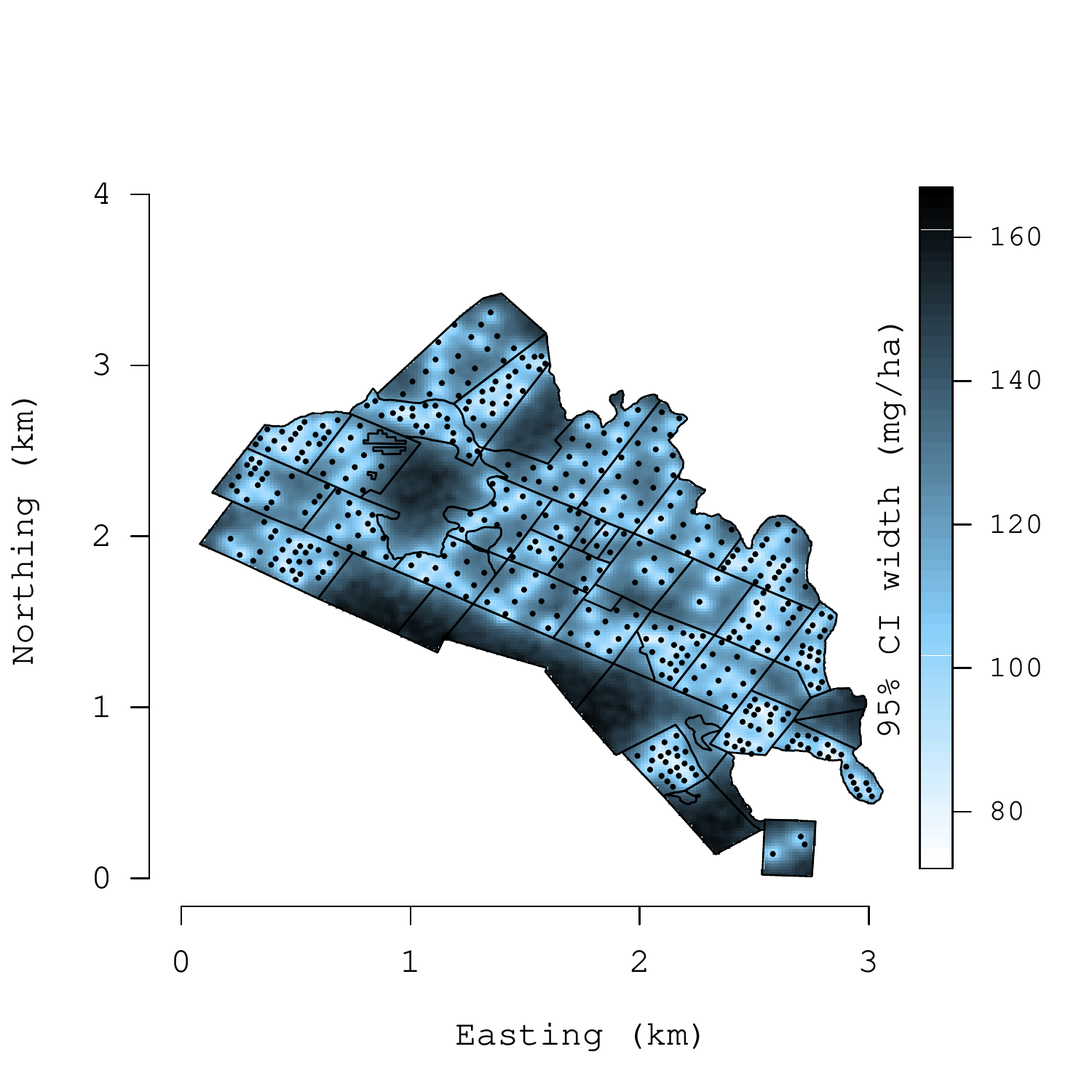}\label{PEF-y-CI-339}}\\
\subfigure[Predicted AGB with $n_u^*=170$, $n_x^*=6$]{\includegraphics[width=8cm]{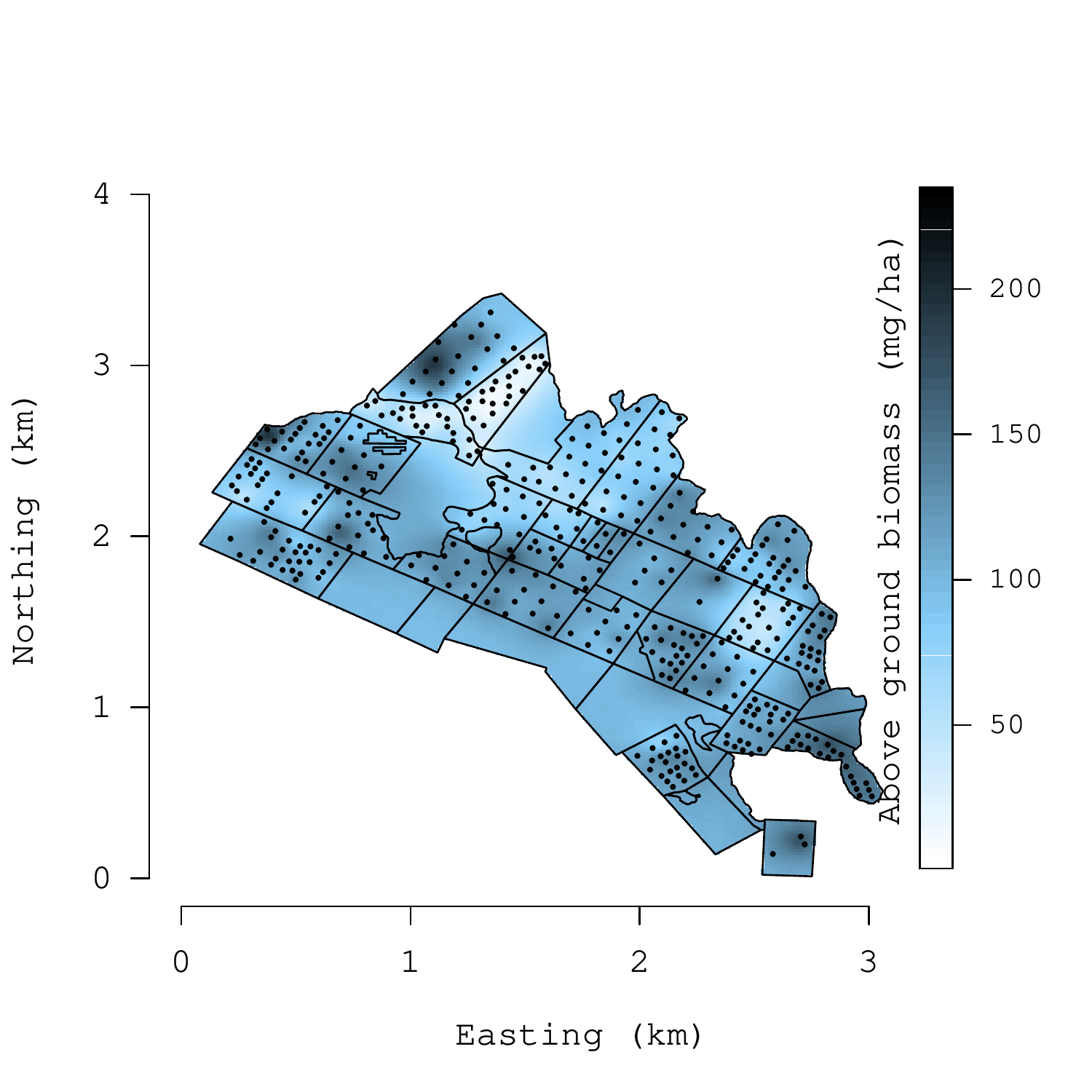}}
\subfigure[95\% prediction interval width for AGB with $n_u^*=170$, $n_x^*=6$]{\includegraphics[width=8cm]{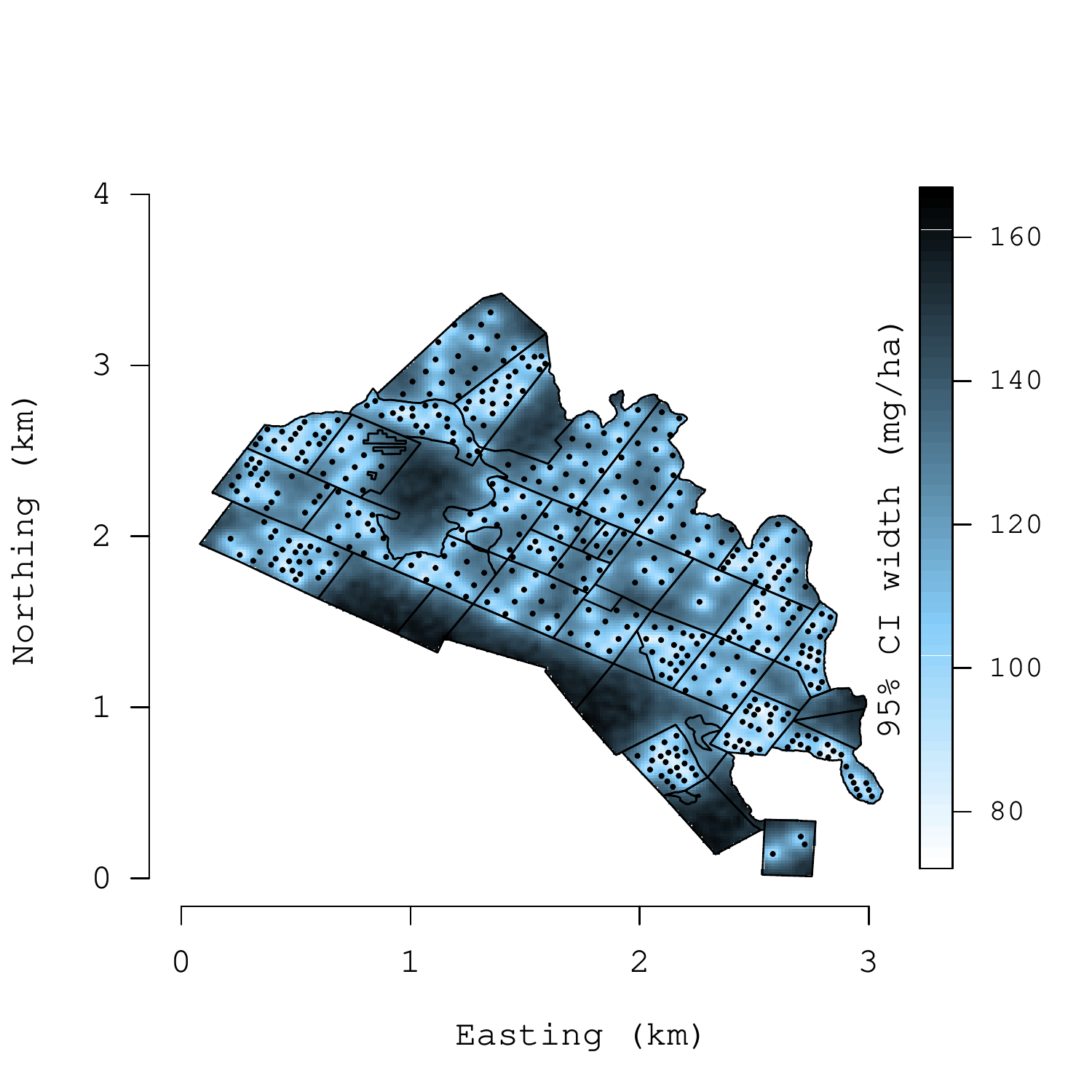}\label{PEF-y-CI-170}}\\
\caption{Posterior predictive distribution's median and width of $95\%$ prediction interval for AGB.}\label{PEF-y}
\end{figure}

Significant $\alpha$ parameter estimates suggest LiDAR signal trends captured by the low-rank models are useful for explaining variability in AGB. The impact of latent LiDAR regressors is seen in Figure~\ref{trans-y} where both models capture AGB trends within MUs and along the example transect. Clearly spatial smoothing occurs---there should likely be more abrupt changes in median AGB across MUs---however there is nothing to inform AGB predictions except for the elevation predictor variable, representation of the LiDAR signals, and residual spatial random effects, all three of which are smoothly varying across the domain. Other candidate models, including those presented in the Supplementary Material, produce similar AGB profiles. We could certainly add a MU indicator or additional location specific predictors to help inform AGB prediction. However, again, these data are rarely available in applied settings and a key objective of this analysis was to assess the usefulness of the latent LiDAR regressors for modeling AGB. Indeed, even lacking additional location specific predictor variables the candidate models yield very useful AGB data products that are critical inputs to forest management and MRV systems. For example, Figure~\ref{PEF-y} offers candidate models' AGB posterior predictive median and associated measure of uncertainty at a 15$\times$15 meter resolution for the entire PEF. This figure shows the candidate models deliver nearly identical AGB prediction and uncertainty maps despite the large reduction in space-height process dimension. As expected, more precise AGB prediction occurs in proximity to observed PSP as shown by narrow 95\% CI intervals in Figures~\ref{PEF-y-CI-339} and \ref{PEF-y-CI-170}.

\clearpage

\section{Summary}\label{Sec: Summary}
We have developed and implemented a class of Bayesian hierarchical models to jointly model LiDAR signals and AGB and effectively exploit the information from the high-dimensional LiDAR signals to explain AGB variability. We account for spatial dependence among and within the high-dimensional LiDAR signals and predict the LiDAR signals and AGB at arbitrary spatial locations and heights. We circumvent computational bottlenecks presented by the LiDAR signal dimensionality and number of spatial locations by applying reduced-rank predictive processes, a collapsed MCMC framework, and some efficient numerical linear algebra.

We opted for a fully process-based approach using covariance functions to exploit the easy constructibility and interpretability of the joint models. Alternative approaches could build upon existing functional data models that treat the high-dimensional signals as a function of space and height. For instance, one could possibly adapt the approach of \cite{yang2015}, who mapped agricultural soil properties, to build joint AGB-LiDAR models. Properties of these models and, in particular, their scalability to massive datasets still need to be explored. 

Substantive contributions from the current PEF analysis include LiDAR-based maps of AGB with associated uncertainty that can inform analyses of MU specific silvicultural experiments and also serve as baseline estimates, with uncertainty, for future management and experiments. More broadly, we believe this modeling framework will be employed for future explorations and analysis relating LiDAR and similar high-dimensional signal data---generated by the missions detailed in Section~\ref{Sec: Intro}---with AGB and other forest variables of interest. Future methodological extensions include analyzing several forest variables (e.g., AGB by tree species or structural variables such as density and basal area) perhaps correlated among themselves, as well as accounting for spatiotemporal associations. There is also considerable interest in adapting the proposed framework to model non-Gaussian forest variables such as forest/non-forest, fire risk categories, and species or functional types. We plan to extend this joint modeling framework to accommodate additional sparsely sampled high-dimensional signal data such as hyper-spectral data that is similar to LiDAR but records information across the electromagnetic spectrum and can provide information on forest species or tree health status.

Our focus was on modeling the space-height structure of LiDAR signals to improve the prediction of $i$) signals at non-sampled locations and $ii$) AGB at locations where the signal may or may not have been observed. If one was interested in modeling vegetation structure, such as leaf area density, then a MacArthur-Horn transformation \citep{macarthur69} could be applied in either a pre-processing step prior to model fitting, or in a posterior predictive fashion (one-for-one using samples from $z(\ell)$'s posterior distribution) to generate posterior distributions of the transformed signals. Using the proposed joint model, future work could test if such signal transformation increases the explained variability in AGB (or similar forest variables of interest), via the $\balpha$ coefficients.

Future analysis of LiDAR and forestry data will need to cope with massive amounts of data and increasing demands on model scalability. Here, we could considerably enhance the scalability of the predictive process using the multi-resolution extensions in \cite{katzfuss2015}, where we construct a sequence of nested predictive processes over a nested partition of the spatial domain. Alternatively, recent developments in massively scalable sparsity-inducing Nearest-Neighbor Gaussian Processes or NNGPs \cite{datta2015} can be exploited. Our framework seamlessly accommodates such processes---we replace $u(\ell)$ and $v(\bs)$ in (\ref{Eq: Full_BHM}) with their NNGP counterparts instead of predictive processes. Rather than dimension reduction, scalability will be achieved exploiting sparsity structures in the resulting precision matrices.

\section*{Acknowledgments}
Andrew Finley was supported by National Science Foundation (NSF) DMS-1513481, EF-1137309, EF-1241874, and EF-1253225, as well as NASA Carbon Monitoring System grants. Sudipto Banerjee was supported by NSF DMS-1513654 and IIS-1562303.

\bibliographystyle{asa}
\bibliography{FBZC15}	
\label{lastpage}

\newpage
\section*{Supplementary Data}

\subsection*{Exploratory data analysis}
\begin{figure}[!htbp]
\includegraphics[width=16cm]{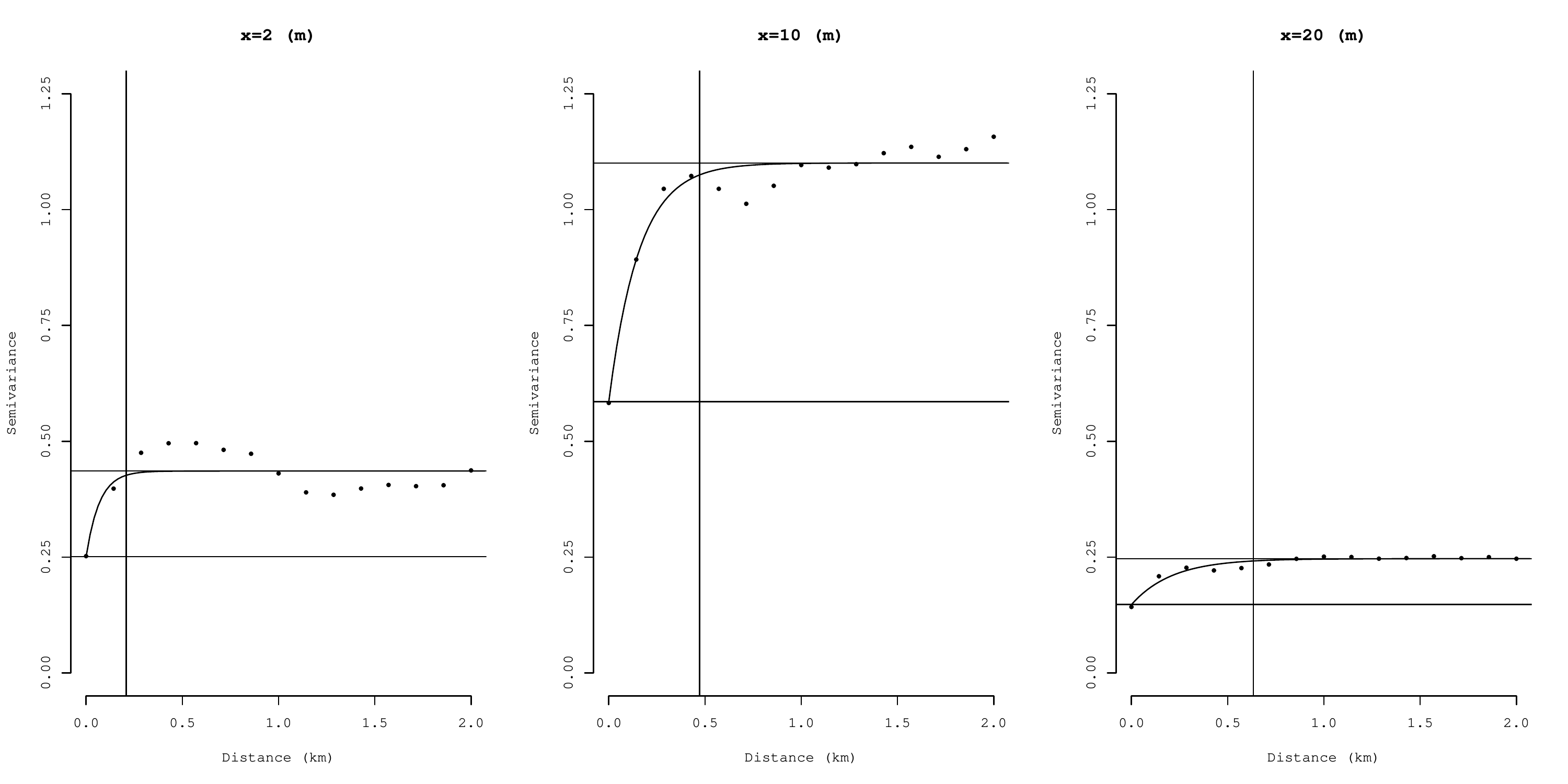}
\caption{Exploratory semivariograms constructed using $\bz$ at the given height $x$ noted in the panel title.}
\end{figure}

\subsection*{Predictive process counterparts for dimension reduction}\label{Sec: Dim_Red}
\noindent To implement the computations necessary for estimating (5) when $n$ is large, we will exploit reduced rank processes to achieve dimension reduction. Such processes usually arise as basis expansions of the original process with fewer number of basis functions than the number of data points. This yields ``low-rank'' processes. Every choice of basis functions yields a process and there are far too many choices to enumerate here; see, e.g., \cite{wikle_2011} for an excellent overview of these methods. Here, we opt for a particularly convenient choice, the predictive process \citep[][]{Banerjee_etall_2008}, which derives the basis functions from taking the conditional expectation of the original process, often called the ``parent'' process, given its realizations over a fixed set of points, often referred to as ``knots.'' These knots are much smaller in number than the original number of points.      

Let $\calS^*_u = \{\mathbf{s}_{u,1}^*,\mathbf{s}_{u,2}^*,\ldots,\mathbf{s}_{u,n_u^*}^*\}$  and $\calS^*_v=\{\mathbf{s}_{v,1}^*,\mathbf{s}_{v,2}^*,\ldots,\mathbf{s}_{v,n^*_v}^*\}$ be two sets of spatial knots to be used for constructing the predictive process counterparts of $u(\ell)$ and $v(\bs)$, respectively. Let $\calX^* = \{x_1^*,x_2^*,\ldots,x_{n_x^*}^*\}$ be a set of knots for heights in the LiDAR signal. We let $\calL^*=\{\ell_1^*,\ell_2^*,\ldots,\ell_{n^*}\}$ be an enumeration of the space-height knots, where each $\ell_i^* = (\bs_{u,j}^*,x_k^*)$ for some $\bs_{u,j}^*\in \calS^*_u$ and $x_k^*\in \calX^*$, and define 
\begin{align}\label{Eq: Predictive_Processes}
\begin{array}{rll}
 \tildeu(\ell) &=& \mbox{E}\left[u(\ell)\given \{u(\ell_i^*)\}\right] = \sum_{i=1}^{n^*}b_{u,i}(\ell) u(\ell_i^*)\; \mbox{ and }\; \\
 \tildev(\bs) &=& \mbox{E}\left[v(\bs)\given \{v(\bs_{v,i}^*)\}\right] = \sum_{i=1}^{n_v^*}b_{v,i}(\bs)v(\bs_{v,i}^*)\; .
\end{array}
\end{align}
The $b_{u,i}(\ell)$'s and $b_{v,i}(\bs)$'s are basis functions derived from the respective conditional expectations in (\ref{Eq: Predictive_Processes}). Dimension reduction is achieved by choosing $n^*$ and $n_v^*$ to be much smaller than $n$ and $n_s$, respectively. Even with data over $n_s$ spatial locations and $n_x$ heights so that $n=n_sn_x$, we need to work only with the $u(\ell_i^*)$ and $v(\bs_i^*)$. Thus, we work with random vectors of dimensions $n^*$ and $n_v^*$ instead of $n$ and $n_s$. If we choose $n^*=n$, $n_v^*=n_v$ and choose their respective knots to coincide with the original points, i.e., $\calS_v^*=\calS$ and $\calL^*=\calL$, then $\tildeu(\ell)$ and $\tildev(\bs)$ coincide with $u(\ell)$ and $v(\bs)$, respectively.

For Gaussian processes, for any $\bs\in \calD$, the $b_{v,i}(\bs)$'s are the solution of $\bC_v^*(\btheta_v)\bb_v(\bs) = \bc_v^*(\bs)$, where $\bb_v(\bs)$ is $n_v^*\times 1$ with $i$-th element $b_{v,i}(\bs)$,  $\bC_v^*(\btheta_v)$ is $n_{v}^*\times n_{v}^*$ with $(i,j)$-th element $C_v(\bs_{v,i}^*, \bs_{v,j}^*;\btheta)$ and $\bc_v^*(\bs)$ is $n_v^*\times 1$ with $i$-th entry $C_v(\bs,\bs_{v,i}^*;\btheta_v)$. Similarly, for any $\ell\in \calD\times \calH$, we solve the $n^*\times n^*$ system $\bC_u^*(\ell)\bb_u(\ell) = \bc_u^*(\ell)$, where $\bb_u(\ell)$ is $n^*\times 1$ with elements $b_{u,i}(\ell)$, $\bC_u^*(\ell)$ is $n^*\times n^*$ with entries $C_{u}(\ell_i^*,\ell_j^*;\btheta_u)$, and $\bc_u^*(\ell)$ is $n^*\times 1$ with entries $C_u(\ell,\ell_i^*)$. 

The predictive process yields the variances of the residual processes $u(\ell)-\tildeu(\ell)$ and $v(\bs)--\tildev(\bs)$ as $\delta^2_u(\ell) = C_u[\ell,\ell] - \bc_u^{*\top}(\ell)\bC_u^{*-1}(\btheta_u)\bc_u^*(\ell)$ and $\delta^2_v(\bs) = C_v(\bs,\bs) - \bc_v^{*\top}(\bs)\bC_v^{*-1}(\btheta_v)\bc_v^*(\bs)$, respectively. To compensate for the smoothing caused by the reduced-rank models, we further construct independent processes $\tildeeps_u(\ell)\stackrel{ind}{\sim} N(0,\delta^2_u(\ell))$ and $\tildeeps_v(\bs)\stackrel{ind}{\sim} N(0,\delta^2_v(\bs))$ and employ $\tildeu(\ell)+\tildeeps_u(\ell)$ and $\tildev(\bs)+\tildeeps_v(\bs)$ for dimension reduction. This adjustment is called a ``bias-adjustment'' as it fixes an over-estimation of the variability at the origin by the reduced-rank processes and provides a better approximation to the parent process \citep{Finley_etall_2009}. 

Replacing the processes in (5) with their predictive process counterparts and introducing the residual adjustments discussed above, produces the following reduced-rank Bayesian hierarchical model
\begin{align}\label{Eq: Predictive_Process_BHM}
 & p(\bTheta) \times N(\bbeta_y\given \bmu_{\beta_y},\bV_{\beta_y})\times N(\bbeta_z\given \bmu_{\beta_z},\bV_{\beta_z}) \times N(\balpha\given \bmu_{\alpha},\bV_{\alpha}) \times N(\bv^*\given \bzero, \bC_v^*(\btheta_v))  \nonumber\\ 
 &\quad\times N(\bu^*\given \bzero, \bC_u^*(\btheta_u)) \times \prod_{j=1}^{n_s} N\left(y(\bs_j)\given \bq_y^{\top}(\bs_j)\bbeta_y + \balpha^{\top}\bB(\bs_j)\bu^* + \bb_{v}^{\top}(\bs_j)\bv^*,d_y^2(\bs_j)\right) \nonumber\\ 
&\qquad\quad \times \prod_{i=1}^{n} N\left(z(\ell_i)\given \bq_z(\ell_i)^{\top}\bbeta_z + \sum_{j=1}^{n^*}b_{u,j}(\ell_i)u(\ell_j^*), d_z^2(\ell_i)\right) \; ,
\end{align}
where $p(\bTheta)$ is as in (6), $d_z^2(\ell_i) = \tau_z^2(x_k) + \delta^2_u(\ell_i)$, $d_y^2(\bs_j) = \tau_y^2 + \sum_{k=1}^{n_x}\alpha_k^2 \delta_u^2(\bs_j,x_k)+\delta^2_v(\bs_j)$, $\balpha$ is as in (5), each $\bB(\bs_j)$ is $n_x\times n^*$ with $(k,i)$-th element $b_{u,i}(\bs_j, x_k)$, $\bu^*$ is $n^*\times 1$ obtained by stacking the $u(\ell_j^*)$'s conformably with $\bB(\bs_j)$, $\bv^*$ is $n_v^*\times 1$ with elements $v(\bs^*_{v,i})$, and $\bC^*_u(\btheta_u)$ and $\bC^*_v(\btheta_v)$ are the covariance matrices for $\bu^*$ and $\bv^*$, respectively. Further savings accrue if we let $\balpha$ in (\ref{Eq: Predictive_Process_BHM}) be $n_x^*\times 1$ with entries $\alpha(x_k^*)$, whence $\bB(\bs_j)$ is $n_x^*\times n^*$ with $(k,i)$-th element $b_{u,i}(\bs_j, x_k^*)$. Letting $\calS_v^*=\calS$ and $\calL^*=\calL$ makes (\ref{Eq: Predictive_Process_BHM}) equal to the full model (5). 

We use a Markov chain Monte Carlo (MCMC) algorithm for generating exact inference from the posterior distribution of all unknown parameters in \ref{Eq: Predictive_Process_BHM}) detailed in Supplemental Material Section~\ref{Sec: Estimation} 

\subsection*{Bayesian estimation}\label{Sec: Estimation}
We use a Markov chain Monte Carlo (MCMC) algorithm for generating exact inference from the posterior distribution of all unknown parameters in (8). To expedite convergence, we use a ``collapsed'' model by integrating out $\bu^*$ and $\bv^*$ from (8). Let $\bB_u(\btheta_u)$ be the $n\times n^*$ matrix with $(i,j)$-th element $b_{u,j}(\ell_i)$, $\bG(\btheta_u,\balpha)$ be $n_s\times n^*$ with rows $\balpha^{\top}\bB(\bs_i)$, $\bB_v(\btheta_v)$ be $n_s\times n_v^*$ with $(i,j)$-th element $b_{v,j}(\bs_{v,i}^*)$, $\bD_y(\btheta_v,\btheta_u,\balpha,\taus_y)$ be the diagonal matrix with elements $\tau_y^2 + \sum_{k=1}^{n_x}\alpha_k^2 \delta_u^2(\bs_i,x_k)+\delta^2_v(\bs_i)$ arranged conformably with $y(\bs_i)$'s in $\by$ and $\bD_z(\btheta_u,\btau_z^2)$ be diagonal with $\tau^2_z(x_k) + \delta_u^2(\ell_i)$ arranged along the diagonal conformably with $z(\ell_i)$'s in $\bz$. Then, $\bV_z = \bB_u(\btheta_u)\bC^*_u(\btheta_u)\bB_u(\btheta_u)^{\top} + \bD_z(\btheta_u,\btau_z^2)$ is the $n\times n$ variance-covariance matrix for $\bz$, $\bV_y = \bG(\btheta_u,\balpha)\bC_u^*(\btheta_u)\bG(\btheta_u,\balpha)^{\top} + \bB_v(\btheta_v)\bC^*_v(\btheta_v)\bB_v(\btheta_v)^{\top} + \bD_y(\btheta_v,\btheta_u,\balpha,\taus_y)$ is the $n_s\times n_s$ variance-covariance matrix for $\by$, and $\bV_{zy} = \bB_u(\btheta_u)\bC^*_u(\btheta_u)\bG^{\top}(\btheta_u,\balpha)$ is the $n\times n_s$ cross-covariance matrix between $\bz$ and $\by$, where $\bz$ is $n\times 1$ with elements $z(\ell_i)$, $\by$ is $n_s\times 1$ with elements $y(\bs_i)$. We write the model in terms of the above matrices as   
\begin{align}\label{Eq: Collapsed_Model}
  &p(\btheta_v)\times p(\btheta_u) \times N(\bbeta_y\given \bmu_{\beta_y},\bV_{\beta_y})\times N(\bbeta_z\given \bmu_{\beta_z},\bV_{\beta_z}) \times N(\balpha\given \bmu_{\alpha},\bV_{\alpha}) \nonumber\\
  &\qquad \times N(\by\given \bQ_y\bbeta_y, \bV_y) \times N\left(\bz\given \bQ_z\bbeta_z + \bV_{zy}\bV_y^{-1}(\by-\bQ_y\bbeta_y), \bV_z -\bV_{zy}\bV_y^{-1}\bV_{zy}^{\top}\right)\; ,
\end{align}
where 
$\bQ_z$ is $n\times p_z$ with rows $\bq_z^{\top}(\ell_i)$ stacked conformably with $\bz$ and $\bQ_y$ is $n_s\times p_y$ with rows $\bq_y^{\top}(\bs_i)$ stacked conformably with $\by$. 


We use random-walk Metropolis steps to update the parameters $\{\btheta_u, \btheta_v, \balpha,\tau^2_y, \btau^2_z\}$ as one block which requires evaluating the multivariate Gaussian likelihoods in (\ref{Eq: Collapsed_Model}) and will benefit from efficient numerical linear algebra for the inverse and determinant of the variance covariance matrices. We can accomplish this effectively using two functions: $\bL = \texttt{chol}(\bM)$ which computes the Cholesky factorization for a positive definite matrix $\bM = \bL\bL^{\top}$, where $\bL$ is lower-triangular, and $\bX = \texttt{trsolve}(\bT,\bB)$ which solves the triangular system $\bT\bX=\bB$ for a triangular (lower or upper) matrix $\bT$. Further details follow.         

The joint density for $\by$ and $\bz$ in (\ref{Eq: Collapsed_Model}) is, in fact, $N(\bw\given \bQ\bbeta, \bA\bJ\bA^{\top} + \bD)$, where $\bw$ is the $(n+n_s)\times 1$ vector obtained by stacking $\bz$ over $\by$, $\bQ$ is block diagonal with blocks $\bQ_z$ and $\bQ_y$, $\bbeta$ is $(p_z+p_y)\times 1$ obtained by stacking $\bbeta_z$ over $\bbeta_y$, $\bA$ is $(n+n_s)\times (n^*+n_v^*)$ partitioned into a $2\times 2$ block matrix with first row $\begin{bmatrix}\bB_u(\btheta_u) : \bO\end{bmatrix}$ and second row $\begin{bmatrix}\bG(\btheta_u,\balpha) : \bB_v(\btheta_v)\end{bmatrix}$, $\bJ$ is block diagonal with blocks $\bC_u^*(\btheta_u)$ and $\bC_v^*(\btheta_v)$, and $\bD$ is block diagonal with blocks $\bD_z(\btheta_u,\btau_z^2)$ and $\bD_y(\btheta_v,\btheta_u,\balpha,\taus_y)$.     
We now compute    
\begin{align}\label{Eq: SWM}
 \left(\bA\bJ\bA^{\top} + \bD\right)^{-1} &= 
 \bD^{-1/2}(\bI - \bH^{\top}\bH)\bD^{-1/2}\;,
\end{align}
where $\bH$ is obtained by first computing $\bW=\bD^{-1/2}\bA$, then the Cholesky factorization $\bL = \texttt{chol}(\bJ^{-1} + \bW^{\top}\bW)$, and finally solving the triangular system $\bH = \texttt{trsolve}(\bL,\bW^{\top})$.  Having obtained $\bH$, we compute $\be=\bw-\bQ\bbeta$, $\bm = \bD^{-1/2}\be$, $\bn = \bH\bm$, and obtain $\bT = \texttt{chol}(\bI_{n^*}-\bH\bH^{\top})$. The log-target density for $\{\btheta_u, \btheta_v, \balpha,\tau^2_y, \tau^2_z\}$ is then computed as
\begin{align}\label{Eq: Metrop_target_density}
& \log p(\btheta_u) + \log p(\btheta_v) + \log p(\tau^2_y) + \sum_{k=1}^{n_x}\log p(\tau_z^2(x_k)) -\frac{1}{2}(\balpha-\bmu_{\alpha})^{\T}\bV_{\alpha}^{-1}(\balpha-\bmu_{\alpha}) \nonumber\\ 
& \qquad- \frac{1}{2}\sum_{i=1}^{n+n_s} d_{ii} + \sum_{i=1}^{n^*+n_v^*} \log t_{ii} - \frac{1}{2}(\bm^{\top}\bm - \bn^{\top}\bn)\; ,
\end{align}
where $d_{ii}$'s and $t_{ii}$'s are the diagonal elements of $\bD$ and $\bT$, respectively. The total number of flops required for evaluating the target is $O((n+n_s)(n^*+n_v^*)^{*3})\approx O(nn^{*3})$ (since $n >> n_s$ and typically we choose $n^* >> n_v^*$) which is considerably cheaper than the $O(n^3)$ flops that would have been required for the analogous computations in (5). In practice, Gaussian proposal distributions are employed for the Metropolis algorithm and all parameters with positive support are transformed to their logarithmic scale. Therefore, the necessary Jacobian adjustments are made to (\ref{Eq: Metrop_target_density}) by adding some scalar quantities which is negligible in terms of computational costs.

Starting with initial values for all parameters, each iteration of the MCMC executes the above calculations to provide a sample for $\{\btheta_u, \btheta_v, \balpha,\tau^2_y, \btau^2_z\}$. The regression parameter $\bbeta$ is then sampled from its full conditional distribution. If $\bV_w = \bA\bJ\bA^{\T} + \bD$ as in (\ref{Eq: SWM}), $\bmu_{\beta}$ is $(p_z+p_y)\times 1$ with $\bmu_{\beta_z}$ stacked over $\bmu_{\beta_y}$ and $\bV_{\beta}$ is block diagonal with blocks $\bV_{\beta_z}$ and $\bV_{\beta_y}$, then the full conditional distribution for $\bbeta$ is $N(\bB\bb, \bB)$, where $\bB^{-1} = \bV_{\beta}^{-1} + \bQ^{\top}\bV_w^{-1}\bQ$ and $\bb = \bV_{\beta}^{-1}\bmu_{\beta} + \bQ^{\T}\bV_w^{-1}\bw$. These are efficiently computed as $[\bx : \bX] = \bD^{-1/2}[\by : \bQ]$, $\tilde{\bX} = \bH\bX$ and setting $\bb = \bV_{\beta}^{-1}\bmu_{\beta} + \bX^{\T}\bx - \tilde{\bX}^{\T}\bH\bx$ and $\bL_{B} = \texttt{chol}(\bV_{\beta}^{-1} + \bX^{\T}\bX-\tilde{\bX}^{\T}\tilde{\bX})$. We then set $\bbeta = \texttt{trsolve}(\bL_B^{\T}, \texttt{trsolve}(\bL_B,\bb)) + \texttt{trsolve}(\bL_B,\tilde{\bz})$, where $\tilde{\bz}$ is a conformable vector of independent $N(0,1)$ variables. 

We repeat the above computations for each iteration of the MCMC algorithm using the current values of the process parameters in $\bV_w$. The algorithm described above will produce, after convergence, posterior samples for $\bOmega = \{\btheta_u,\balpha,\btheta_v,\tau^2_y, \btau^2_z, \bbeta_y, \bbeta_z\}$. We can subsequently obtain the posterior samples for $\bu^*$ and $\bv^*$ using exact sampling. To be precise, if $\bg$ is the $(n^* + n_v^*)\times 1$ vector with $\bu^*$ stacked over $\bv^*$, then we seek samples from its posterior predictive distribution  
\begin{align}\label{Eq: Posterior_latent_effects}
 p(\bg \given \by, \bz) = \int p(\bg \given \bOmega, \by,\bz)p(\bOmega \given \by, \bz)d\bOmega\; , 
\end{align}
where $p(\bg\given \bOmega, \by, \bz)$ is $N(\bB\bb, \bB)$, where $\bB = (\bJ^{-1} + \bA^{\top}\bD^{-1}\bA)^{-1}$ and $\bb = \bA^{\top}\bD^{-1}(\bw - \bQ\bbeta)$. Since $n^*+n_v^*$ is chosen to be much smaller than $n+n_s$, obtaining $\texttt{chol}(\bB)$ is not as expensive, but can produce numerical instabilities due to the inverses of $\bC_u^*(\btheta_u)$ and $\bC_v^*(\btheta_v)$ appearing in $\bJ^{-1}$ which we seek to avoid. We execute a numerically stable algorithm exploiting the fact that $\bB = \bK - \bK(\bJ + \bK)^{-1}\bK$, where $\bK^{-1} = \bA^{\top}\bD^{-1}\bA$. 
For each posterior sample of $\bOmega$, we compute $\bL = \texttt{chol}(\bJ + \bK)$, $\bW = \texttt{trsolve}(\bL,\bK)$ and $\bL_{B} = \bK - \bW^{\top}\bW$. We generate an $(n^*+n_v^*)\times 1$ vector $\bz^*\sim N(\bzero,\bI_{n^*+n_v^*})$ and set $\bg = \bL_B(\bz^* + \bL_B^{\top}\bb)$. Repeating this for every posterior sample of $\bOmega$, produces the posterior predictive samples for $\bg$ from (\ref{Eq: Posterior_latent_effects}) and, hence, those for $\bu^*$ and $\bv^*$.

\subsection*{Bayesian prediction}\label{sec: Bayesian implementation}
The following posterior predictive distributions provide predictive inference for $z(\ell_0)$ at any arbitrary space-height coordinate $\ell_0$ and for $y(\bs_0)$ at any arbitrary spatial location $\bs_0$:
\begin{align}\label{Eq: Posterior_Predictive_Sampling}
z(\ell_0) &\sim N\left(\bq_z^{\top}(\ell_0)\bbeta_z + \sum_{j=1}^{n^*}b_{u,j}(\ell_0)u(\ell_j^*), d_z^2(\ell_0)\right)\; \mbox{ and }\; \\ y(\bs_0) &\sim N\left(\bq_y^{\top}(\bs_0)\bbeta_y + \balpha^{\top}\bB(\bs_0)\bu^* + \bb_{v}^{\top}(\bs_0)\bv^*, d_y^2(\bs_0)\right)\; .  
\end{align}
Given predictor variables in $\bq_z^{\top}(\ell_0)$ and $\bq_y^{\top}(\bs_0)$ and drawing from (\ref{Eq: Posterior_Predictive_Sampling}) for each posterior sample of $\bTheta$, $\balpha$, $\bbeta_y$, $\bbeta_z$, $\bv^*$, and $\bu^*$ yields the corresponding posterior predictive sample for $z(\ell_0)$ and $y(\bs_0)$. Posterior predictive samples of the latent processes can also be easily computed as $u(\ell_0) = \sum_{j=1}^{n^*}b_{u,j}(\ell_0)u(\ell_j^*)$ and $v(\bs_0) = \bb_{v}^{\top}(\bs_0)\bv^*$ for each posterior sample of the $u(\ell_j^*)$'s, $\bv^*$ and the process parameters present in the basis functions $b_{u,j}(\ell_0)$ and $\bb_{v}^{\top}(\bs_0)$. 

\subsection*{Approximately optimal selection of knots}
Specifying the number and location of knots is key to dimension reduction in both $u$ and $v$. \cite{Tokdar2011} and \cite{Guhaniyogi2011} discuss various approaches to knot selection approaches for reduced-rank models. Here, we assume the number of spatial knots for $u$ and $v$ are set based on acceptable computing time and placement is achieved using a sequential search algorithm over a fine grid of candidate locations, see \cite{Finley_etall_2009} for details. When the number of candidate locations for height knots is small, we can use an exhaustive search instead of a sequential search. For example, in the subsequent PEF data analysis we coarsen the LiDAR signals by approximately half and consider only $n_x=39$ candidate locations for the height knots; hence, an exhaustive search over all subsets in $\{x_1,x_2,\ldots,x_{n_x}\}$ was computationally feasible. Specifically, height knots $x^*$ are chosen by minimizing
\begin{align*}
\sum_{k=1}^{n_x} \Big(C_u((\bs, x_k), (\bs, x_k))-\bc_{u,s}^\top(x_k,\mathcal{X}^*)\bC_{u,s}^{-1}(\mathcal{X}^*,\mathcal{X}^*)\bc_{u,s}(x_k,\mathcal{X}^*)\Big),
\end{align*}
where $\bc_{u,s}^\top(x_k,\mathcal{X}^*)$ is $n_x^*\times 1$ with $i$-th element $C_u((\bs,x_k),(\bs,x_i^*))$ and $\bC_{u,s}(\mathcal{X}^*,\mathcal{X}^*)$ is $n_x^*\times n_x^*$ with $(i,j)$-th element $C_u((\bs,x_i^*),(\bs,x_j^*))$. Then given the height knot locations, spatial knots $\mathcal{S}_u^*$ for $u$ are chosen by minimizing 
\begin{align*}
\sum_{j=1}^{n_s}\sum_{k=1}^{n_x^*}\Big(C_u((\bs_j,x_k^*),(\bs_j,x_k^*))-\bc_u^{*\top}(\bs_j,x_k^*)\bC_u^{*-1}(\btheta_u)\bc_u^*(\bs_j,x_k^*)\Big).
\end{align*}
Finally, spatial knots $\mathcal{S}_v^*$ for $v$ are selected by minimizing $\sum_{i=1}^{n_s} \delta_v^2(\mathbf s_i)$. Here, $\bc_u^*(\bs_j,x_k^*)$, $\bC_u^*(\btheta_u)$ and $\delta_v^2(\mathbf s_i)$ were defined in Section~\ref{Sec: Dim_Red} in the paper. 

\subsection*{Additional candidate model results for the PEF analysis}
Here we present the results for the intercept only models and those that exclude the spatial process $v(\bs)$. 

\begin{sidewaystable}[!htbp]
\caption {Parameter credible intervals, $50\%\, (2.5\%,97.5\%)$, and goodness-of-fit for the intercept only $n_u^*=339$ and $n_v^*=339$ models. Bold parameter values indicated values that differ from zero where appropriate and bold goodness-of-fit metrics indicate \emph{best} fit.}\label{est-pred-realData-339-with-v}
\small
\begin{tabular}{ccccccc}
\toprule
{Parameter}  & \multicolumn{6}{c}{Height knot models} \\
\cmidrule{2-7}
{}  & {$n_x^*=2$} & {$n_x^*=3$} & {$n_x^*=4$}&{$n_x^*=5$}&{$n_x^*=6$}&{$n_x^*=7$} \\
\midrule
$\beta_y$&1.04(0.87,1.23)&1.07(0.78,1.39)&1.04(0.9,1.2)&1.05(0.88,1.23)&1.05(0.86,1.24)&1.04(0.89,1.2)\\
$\alpha_1$&\textbf{-0.08(-0.14,-0.01)}&-0.09(-0.2,0.01)&\textbf{-0.04(-0.09,0)}&0.02(-0.04,0.07)&-0.05(-0.12,0.01)&-0.05(-0.11,0.01)\\
$\alpha_2$&\textbf{0.32(0.27,0.38)}&0.01(-0.19,0.22)&\textbf{-0.16(-0.23,-0.1)}&-0.05(-0.13,0.03)&\textbf{-0.12(-0.19,-0.04)}&\textbf{-0.08(-0.17,-0.01)}\\
$\alpha_3$&\qquad -&\textbf{0.42(0.31,0.53)}&\textbf{0.11(0.06,0.18)}&0.06(-0.02,0.13)&\textbf{-0.11(-0.18,-0.04)}&\textbf{-0.1(-0.18,-0.04)}\\
$\alpha_4$&\qquad -&\qquad -&\textbf{0.26(0.14,0.36)}&\textbf{0.2(0.13,0.26)}&0.03(-0.04,0.11)&-0.02(-0.1,0.05)\\
$\alpha_5$&\qquad -&\qquad -&\qquad -&\textbf{0.32(0.16,0.45)}&\textbf{0.11(0.03,0.19)}&0.06(-0.02,0.14)\\
$\alpha_6$&\qquad -&\qquad -&\qquad -&\qquad -&\textbf{0.2(0,0.39)}&\textbf{0.16(0.04,0.27)}\\
$\alpha_7$&\qquad -&\qquad -&\qquad -&\qquad -&\qquad -&0.03(-0.19,0.21)\\
$\sigma^2_u$&0.11(0.1,0.11)&0.15(0.14,0.17)&0.48(0.45,0.52)&0.62(0.58,0.67)&0.99(0.9,1.1)&1.18(1.09,1.29)\\
$a$&1.23(1.09,1.36)&1.41(1.26,1.55)&0.78(0.73,0.83)&1.13(1.07,1.2)&0.97(0.91,1.03)&1.04(0.99,1.09)\\
$\gamma$&1(1,1)&0.99(0.98,1)&0.99(0.97,1)&0.99(0.98,0.99)&1(1,1)&0.99(0.99,1)\\
$c$&17.73(15.28,21.53)&11.61(10.46,12.88)&8.98(8.3,9.8)&8.93(8.36,9.54)&7.72(7.18,8.33)&8.31(7.78,8.88)\\
$\sigma^2_v$&0.1(0.07,0.15)&0.15(0.1,0.22)&0.07(0.05,0.1)&0.08(0.05,0.11)&0.09(0.06,0.12)&0.08(0.06,0.11)\\
$\phi_v$&3.47(1.9,5.93)&2.01(1.4,4.75)&3.67(2.41,5.35)&4.24(2.63,5.57)&3.42(1.96,4.02)&5.02(2.36,7.99)\\
$\tau_y^2$&0.03(0.02,0.04)&0.03(0.02,0.04)&0.03(0.02,0.04)&0.03(0.02,0.04)&0.03(0.02,0.04)&0.02(0.02,0.04)\\
 \midrule
 $p_D$&72.83&75.77&66.55&64.72&66.36&62.87\\
 DIC&25394.17&24108.86&19561.92&15282.15&12766.62&\textbf{9669.22}\\
 G&10492.42&8488.55&7009.5&4549.5&3724.98&3251.79\\
 P&10864.7&8967.3&7714.91&5122.17&4284.75&3623.49\\
 D=G+P&21357.12&17455.86&14724.41&9671.68&8009.73&\textbf{6875.27}\\
\bottomrule
\end{tabular} 
\end{sidewaystable}

\begin{sidewaystable}[!htbp]
\caption {Parameter credible intervals, $50\%\, (2.5\%,97.5\%)$, and goodness-of-fit for the intercept only $n_u^*=170$ and $n_v^*=339$ models. Bold parameter values indicated values that differ from zero where appropriate and bold goodness-of-fit metrics indicate \emph{best} fit.}\label{est-pred-realData-170-with-v}
\small
\begin{tabular}{ccccccc}
\toprule
{Parameter}  & \multicolumn{6}{c}{Height knot models} \\
\cmidrule{2-7}
{}  & {$n_x^*=2$} & {$n_x^*=3$} & {$n_x^*=4$}&{$n_x^*=5$}&{$n_x^*=6$}&{$n_x^*=7$} \\
\midrule
$\beta_y$&1.03(0.81,1.25)&1.04(0.71,1.39)&1.04(0.9,1.19)&1.05(0.85,1.25)&1.06(0.81,1.31)&1.05(0.78,1.34)\\
$\alpha_1$&\textbf{-0.22(-0.43,-0.02)}&\textbf{-0.46(-0.67,-0.19)}&-0.03(-0.15,0.1)&-0.04(-0.17,0.12)&-0.13(-0.28,0.03)&-0.11(-0.28,0.08)\\
$\alpha_2$&\textbf{0.73(0.55,0.9)}&\textbf{-0.54(-0.85,-0.1)}&\textbf{-0.26(-0.46,-0.07)}&\textbf{-0.23(-0.4,-0.05)}&-0.12(-0.29,0.03)&-0.1(-0.24,0.04)\\
$\alpha_3$&\qquad -&\textbf{0.59(0.38,0.83)}&\textbf{0.24(0.06,0.42)}&-0.09(-0.26,0.09)&\textbf{-0.26(-0.42,-0.12)}&\textbf{-0.21(-0.37,-0.02)}\\
$\alpha_4$&\qquad -&\qquad -&\textbf{0.28(0.08,0.5)}&\textbf{0.26(0.12,0.44)}&0.09(-0.09,0.28)&-0.08(-0.23,0.06)\\
$\alpha_5$&\qquad -&\qquad -&\qquad -&0.12(-0.2,0.42)&0.12(-0.08,0.3)&0.16(-0.01,0.38)\\
$\alpha_6$&\qquad -&\qquad -&\qquad -&\qquad -&0.13(-0.31,0.47)&0.06(-0.13,0.29)\\
$\alpha_7$&\qquad -&\qquad -&\qquad -&\qquad -&\qquad -&0.13(-0.2,0.5)\\
$\sigma^2_u$&0.05(0.05,0.06)&0.08(0.07,0.09)&0.26(0.23,0.28)&0.37(0.33,0.41)&0.41(0.37,0.46)&0.44(0.4,0.5)\\
$a$&0.73(0.58,0.91)&1.08(0.92,1.32)&0.85(0.77,0.94)&1.21(1.11,1.34)&1.24(1.13,1.37)&1.51(1.37,1.66)\\
$\gamma$&0.99(0.98,1)&0.99(0.94,1)&1(0.99,1)&1(0.99,1)&0.99(0.98,0.99)&0.99(0.98,1)\\
$c$&6.75(6.06,7.49)&5.13(4.53,5.8)&3.36(3.03,3.73)&3.1(2.78,3.43)&2.82(2.54,3.14)&3(2.68,3.34)\\
$\sigma^2_v$&0.09(0.06,0.16)&0.11(0.05,0.22)&0.05(0.04,0.08)&0.06(0.04,0.1)&0.07(0.05,0.11)&0.07(0.05,0.16)\\
$\phi_v$&2.42(1.43,4.99)&1.59(0.94,4.68)&4.65(3.03,7.33)&3.5(2.01,5.89)&2.95(1.46,6.01)&2.67(1.23,6.72)\\
$\tau_y^2$&0.03(0.02,0.05)&0.03(0.01,0.04)&0.03(0.02,0.05)&0.03(0.02,0.04)&0.03(0.02,0.05)&0.03(0.02,0.05)\\
\midrule
$p_D$&79.27&78.95&73.86&68.84&70.81&72.06\\
DIC&27128.67&26426.63&24557.72&22929.39&22393.55&\textbf{21962.06}\\
G&11022.24&9777.49&8133.86&6743.01&6476.55&6154.11\\
P&11298.15&10112.78&8516.51&7207.75&6908.56&6554.33\\
D=G+P&22320.39&19890.27&16650.37&13950.76&13385.11&\textbf{12708.44}\\
\bottomrule
\end{tabular} 
\end{sidewaystable}

\begin{sidewaystable}[!htbp]
\caption {Prediction metrics for the intercept only $n_u^*=339$ and $n_v^*=339$ models. Bold values indicate \emph{best} predictive performance.}\label{est-pred-realData-339-with-v-2}
\small
\begin{tabular}{ccccccc}
\toprule
{Parameter}  & \multicolumn{6}{c}{Height knot models} \\
\cmidrule{2-7}
{}  & {$n_x^*=2$} & {$n_x^*=3$} & {$n_x^*=4$}&{$n_x^*=5$}&{$n_x^*=6$}&{$n_x^*=7$} \\
\midrule
RMSPE for AGB \& LiDAR&0.89&0.855&0.833&0.784&0.775&\textbf{0.771}\\
CRPS for AGB \& LiDAR&1972.57&1908.42&1863.4&\textbf{1772.23}&1778.66&1789.73\\
GRS for AGB \& LiDAR&-2129.19&-1589.73&-1500.15&\textbf{-1168.25}&-1261.61&-1432.38\\
95\% prediction coverage for AGB \&LiDAR&93&93.3&93.7&93.6&94.5&95.1\\ 
\midrule
RMSPE for $\mbox{AGB}\given \mbox{observed LiDAR}$ &0.324&0.319&0.317&\textbf{0.316}&0.317&0.317\\
CRPS for $\mbox{AGB}\given \mbox{observed LiDAR}$ &20.57&20.05&19.96&\textbf{19.87}&19.9&19.9\\
GRS for $\mbox{AGB}\given \mbox{observed LiDAR}$ &129.64&142.56&145.1&146.74&\textbf{146.85}&146.57\\
95\% prediction interval coverage for $\mbox{AGB}\given \mbox{observed LiDAR}$ &89.3&92.9&93.8&94.6&94.6&93.8\\
95\% prediction interval width for $\mbox{AGB}\given \mbox{observed LiDAR}$ &1.01&1.09&1.14&1.27&1.23&1.21\\
\bottomrule
\end{tabular} 
\end{sidewaystable}

\begin{sidewaystable}[!htbp]
\caption {Prediction metrics for the intercept only $n_u^*=170$ and $n_v^*=339$ models. Bold values indicate \emph{best} predictive performance.}\label{est-pred-realData-170-with-v-2}
\small
\begin{tabular}{ccccccc}
\toprule
{Parameter}  & \multicolumn{6}{c}{Height knot models} \\
\cmidrule{2-7}
{}  & {$n_x^*=2$} & {$n_x^*=3$} & {$n_x^*=4$}&{$n_x^*=5$}&{$n_x^*=6$}&{$n_x^*=7$} \\
\midrule
RMSPE for AGB \& LiDAR&0.887&0.859&0.826&0.8&0.797&\textbf{0.793}\\
CRPS for AGB \& LiDAR&1955.91&1910.43&1833.09&1776.21&1767.83&\textbf{1757.67}\\
GRS for AGB \& LiDAR&-1234.03&-1289.35&\textbf{-1171.55}&-1237.36&-1239.89&-1322.01\\
95\% prediction coverage for AGB \& LiDAR&93.9&93.5&92.7&91.5&91.1&90.9\\
 \midrule
 RMSPE for $\mbox{AGB}\given \mbox{observed LiDAR}$ &0.317&0.316&\textbf{0.308}&0.311&0.309&0.31\\
 CRPS for $\mbox{AGB}\given \mbox{observed LiDAR}$ &19.86&19.77&\textbf{19.36}&19.49&\textbf{19.36}&19.37\\
 GRS for $\mbox{AGB}\given \mbox{observed LiDAR}$ &143.19&145.11&150.07&147.28&149.98&\textbf{150.29}\\
 95\% prediction interval coverage for $\mbox{AGB}\given \mbox{observed LiDAR}$ &92&92&91.1&92&92&92.9\\
 95\% prediction interval width for $\mbox{AGB}\given \mbox{observed LiDAR}$ &1.07&1.1&1.07&1.06&1.07&1.08\\
\bottomrule
\end{tabular} 
\end{sidewaystable}

\begin{sidewaystable}[!htbp]
\caption {Parameter credible intervals, $50\%\, (2.5\%,97.5\%)$, and goodness-of-fit for the intercept only models with $n_u^*=339$ and $v(\bs)=0$. Bold parameter values indicated values that differ from zero where appropriate and bold goodness-of-fit metrics indicate \emph{best} fit.}\label{est-pred-realData-339-wo-v}
\small
\begin{tabular}{ccccccc}
\toprule
{Parameter}  & \multicolumn{6}{c}{Height knot models} \\
\cmidrule{2-7}
{}  & {$n_x^*=2$} & {$n_x^*=3$} & {$n_x^*=4$}&{$n_x^*=5$}&{$n_x^*=6$}&{$n_x^*=7$} \\
\midrule
$\beta_y$&1(0.96,1.05)&1(0.93,1.06)&1.02(0.97,1.07)&1.03(0.91,1.16)&1.02(0.93,1.12)&1.02(0.9,1.15)\\
$\alpha_1$&-0.08(-0.16,0.01)&-0.08(-0.25,0.06)&0.01(-0.05,0.06)&\textbf{0.1(0.02,0.17)}&-0.06(-0.14,0.02)&-0.06(-0.14,0.01)\\
$\alpha_2$&\textbf{0.44(0.36,0.53)}&-0.01(-0.29,0.26)&\textbf{-0.22(-0.29,-0.15)}&-0.07(-0.16,0.01)&\textbf{-0.14(-0.22,-0.06)}&\textbf{-0.13(-0.22,-0.04)}\\
$\alpha_3$&\qquad -&\textbf{0.55(0.37,0.72)}&\textbf{0.23(0.15,0.29)}&0.06(-0.02,0.14)&\textbf{-0.2(-0.27,-0.12)}&\textbf{-0.17(-0.24,-0.09)}\\
$\alpha_4$&\qquad -&\qquad -&0.1(-0.05,0.23)&\textbf{0.31(0.25,0.39)}&\textbf{0.1(0.02,0.19)}&\textbf{-0.09(-0.17,0)}\\
$\alpha_5$&\qquad -&\qquad -&\qquad -&0.17(-0.03,0.37)&0.08(-0.03,0.18)&\textbf{0.18(0.08,0.26)}\\
$\alpha_6$&\qquad -&\qquad -&\qquad -&\qquad -&0.1(-0.08,0.25)&0.02(-0.12,0.16)\\
$\alpha_7$&\qquad -&\qquad -&\qquad -&\qquad -&\qquad -&-0.06(-0.35,0.3)\\
$\sigma^2_u$&0.11(0.1,0.12)&0.16(0.14,0.17)&0.48(0.44,0.52)&0.61(0.57,0.66)&0.98(0.89,1.07)&1.18(1.09,1.29)\\
$a$&1.36(1.14,1.55)&1.42(1.24,1.58)&0.77(0.72,0.83)&1.12(1.06,1.19)&0.97(0.92,1.03)&1.04(0.99,1.1)\\
$\gamma$&1(0.99,1)&0.99(0.98,1)&0.99(0.99,1)&0.99(0.98,1)&1(1,1)&1(1,1)\\
$c$&16.93(14.94,19.32)&11.85(10.58,12.98)&8.98(8.24,9.78)&8.87(8.25,9.55)&7.8(7.14,8.41)&8.32(7.85,8.88)\\
$\tau_y^2$&0.12(0.11,0.15)&0.14(0.12,0.17)&0.1(0.09,0.12)&0.11(0.09,0.12)&0.1(0.08,0.11)&0.1(0.09,0.11)\\
 \midrule
 $p_D$&69.59&69.9&67.32&66.38&65.26&60.4\\
 DIC&25595.71&24341.16&19718.84&15440.32&12865.25&\textbf{9833.96}\\
 G&10480.05&8512.66&7007.11&4594.12&3789.84&3285.28\\
 P&10847.3&9084&7474.24&4978.46&4383.57&3770.82\\
 D=G+P&21327.35&17596.66&14481.36&9572.58&8173.41&\textbf{7056.11}\\
\bottomrule
\end{tabular} 
\end{sidewaystable}

\begin{sidewaystable}[!htbp]
\caption {Parameter credible intervals, $50\%\, (2.5\%,97.5\%)$, and goodness-of-fit for the intercept only models with $n_u^*=170$ and $v(\bs)=0$. Bold parameter values indicated values that differ from zero where appropriate and bold goodness-of-fit metrics indicate \emph{best} fit.}\label{est-pred-realData-170-wo-v}
\small
\begin{tabular}{ccccccc}
\toprule
{Parameter}  & \multicolumn{6}{c}{Height knot models} \\
\cmidrule{2-7}
{}  & {$n_x^*=2$} & {$n_x^*=3$} & {$n_x^*=4$}&{$n_x^*=5$}&{$n_x^*=6$}&{$n_x^*=7$} \\
\midrule
$\beta_y$&1.01(0.95,1.07)&1.1(0.88,1.33)&1.05(0.91,1.17)&1.05(0.86,1.21)&1.06(0.81,1.34)&1.12(0.71,1.56)\\
$\alpha_1$&\textbf{-0.26(-0.45,-0.08)}&\textbf{-1.05(-1.2,-0.89)}&\textbf{0.11(0.01,0.22)}&\textbf{0.16(0.01,0.29)}&\textbf{-0.24(-0.39,-0.11)}&\textbf{-0.51(-0.62,-0.4)}\\
$\alpha_2$&\textbf{0.85(0.68,1.02)}&\textbf{-1.33(-1.54,-1.14)}&\textbf{-0.19(-0.35,-0.02)}&\textbf{-0.15(-0.29,0)}&\textbf{-0.17(-0.28,-0.04)}&0.05(-0.13,0.22)\\
$\alpha_3$&\qquad -&\textbf{0.2(0,0.4)}&\textbf{0.45(0.29,0.62)}&0.04(-0.12,0.19)&\textbf{-0.45(-0.56,-0.34)}&\textbf{-0.62(-0.73,-0.51)}\\
$\alpha_4$&\qquad -&\qquad -&0.08(-0.11,0.26)&\textbf{0.43(0.31,0.55)}&\textbf{0.17(0.03,0.3)}&-0.06(-0.2,0.09)\\
$\alpha_5$&\qquad -&\qquad -&\qquad -&-0.05(-0.28,0.22)&-0.09(-0.27,0.05)&-0.04(-0.18,0.11)\\
$\alpha_6$&\qquad -&\qquad -&\qquad -&\qquad -&0.11(-0.13,0.48)&0.06(-0.14,0.26)\\
$\alpha_7$&\qquad -&\qquad -&\qquad -&\qquad -&\qquad -&-0.2(-0.5,0.15)\\
$\sigma^2_u$&0.05(0.05,0.06)&0.07(0.06,0.08)&0.25(0.23,0.28)&0.36(0.32,0.39)&0.41(0.37,0.46)&0.44(0.39,0.49)\\$a$&0.84(0.67,1.04)&1.25(1.01,1.51)&0.86(0.77,0.95)&1.21(1.11,1.34)&1.26(1.15,1.37)&1.46(1.33,1.58)\\
$\gamma$&0.99(0.95,1)&0.98(0.92,0.99)&0.99(0.98,0.99)&1(0.99,1)&0.99(0.98,1)&1(0.99,1)\\
$c$&6.66(5.95,7.37)&4.89(4.25,5.5)&3.31(2.97,3.7)&3.03(2.74,3.4)&2.94(2.67,3.25)&2.91(2.58,3.21)\\
$\tau_y^2$&0.11(0.09,0.13)&0.02(0.01,0.03)&0.08(0.06,0.09)&0.08(0.07,0.09)&0.06(0.04,0.07)&0.02(0.01,0.03)\\
 \midrule
 $p_D$&77.92&77.52&78.26&74.89&71.14&69.16\\
 DIC&27294.97&26509.74&24654.31&23029.21&22431.91&\textbf{21991.62}\\
 G&10981.82&9829.4&8164.04&6802.96&6470.46&6236.85\\
 P&11216.56&10023.22&8581.87&7133.01&6915.99&6647.26\\
 D=G+P&22198.37&19852.62&16745.91&13935.97&13386.45&\textbf{12884.12}\\
\bottomrule
\end{tabular} 
\end{sidewaystable}

\begin{sidewaystable}[!htbp]
\caption {Prediction metrics for the intercept only models with $n_u^*=339$ and $v(\bs)=0$. Bold values indicate improved prediction performance.}\label{est-pred-realData-339-wo-v-2}
\small
\begin{tabular}{ccccccc}
\toprule
{Parameter}  & \multicolumn{6}{c}{Height knot models} \\
\cmidrule{2-7}
{}  & {$n_x^*=2$} & {$n_x^*=3$} & {$n_x^*=4$}&{$n_x^*=5$}&{$n_x^*=6$}&{$n_x^*=7$} \\
\midrule
RMSPE for AGB \&LiDAR&0.889&0.856&0.832&0.785&0.777&\textbf{0.772}\\
CRPS for AGB \& LiDAR&1975.14&1914.2&1863.92&\textbf{1775.47}&1782.35&1793.04\\
GRS for AGB \& LiDAR&-2151.04&-1645.56&-1524.12&\textbf{-1201.28}&-1288.95&-1468.77\\
95\% prediction coverage for AGB \& LiDAR&93.5&93.9&94.2&94.1&95&95.8\\ 
\midrule
RMSPE for $\mbox{AGB}\given \mbox{observed LiDAR}$ &0.393&0.399&0.363&0.363&\textbf{0.355}&0.36\\
CRPS for $\mbox{AGB}\given \mbox{observed LiDAR}$ &25.08&25.48&23.19&23.24&\textbf{22.61}&23.07\\
GRS for $\mbox{AGB}\given \mbox{observed LiDAR}$ &97.24&92.99&115.05&112.58&\textbf{118.49}&111.01\\
95\% prediction interval coverage for $\mbox{AGB}\given \mbox{observed LiDAR}$ &96.4&98.2&95.5&97.3&96.4&97.3\\
95\% prediction interval width for $\mbox{AGB}\given \mbox{observed LiDAR}$ &1.48&1.66&1.42&1.57&1.5&1.66\\
\bottomrule
\end{tabular} 
\end{sidewaystable}

\begin{sidewaystable}[!htbp]
\caption {Prediction metrics for the intercept only models with $n_u^*=170$ and $v(\bs)=0$. Bold values indicate improved prediction performance.}\label{est-pred-realData-170-wo-v-2}
\small
\begin{tabular}{ccccccc}
\toprule
{Parameter}  & \multicolumn{6}{c}{Height knot models} \\
\cmidrule{2-7}
{}  & {$n_x^*=2$} & {$n_x^*=3$} & {$n_x^*=4$}&{$n_x^*=5$}&{$n_x^*=6$}&{$n_x^*=7$} \\
\midrule
RMSPE for AGB \& LiDAR&0.887&0.862&0.827&0.802&\textbf{0.796}&\textbf{0.796}\\
CRPS for AGB \& LiDAR&1956.61&1915.61&1836.23&1781.14&1768.94&\textbf{1763}\\
GRS for AGB \& LiDAR&-1284.79&-1264.35&\textbf{-1173.59}&-1225.8&-1285.01&-1321.05\\
95\% prediction coverage for AGB \& LiDAR&94.6&94.1&92.9&91.8&91.5&91.3\\
\midrule
RMSPE for $\mbox{AGB}\given \mbox{observed LiDAR}$ &0.362&\textbf{0.31}&0.325&0.325&0.315&0.315\\
CRPS for $\mbox{AGB}\given \mbox{observed LiDAR}$ &23.14&\textbf{19.49}&20.8&20.81&19.88&19.89\\
GRS for $\mbox{AGB}\given \mbox{observed LiDAR}$ &115.94&\textbf{148.24}&138.97&139.14&146.48&146\\
95\% prediction interval coverage for $\mbox{AGB}\given \mbox{observed LiDAR}$ &96.4&97.3&95.5&95.5&94.6&95.5\\
95\% prediction interval width for $\mbox{AGB}\given \mbox{observed LiDAR}$ &1.41&1.28&1.23&1.26&1.21&1.26\\
\bottomrule
\end{tabular} 
\end{sidewaystable}


\end{document}